\documentclass[a4paper]{article}
\usepackage{a4wide}
\usepackage{titling}
\usepackage[hyphenbreaks]{breakurl}
\usepackage[hyphens]{url}
\usepackage{authblk}
\usepackage{seqsplit}
\usepackage{graphicx}

\usepackage{booktabs}
\usepackage{xcolor}
\usepackage{soul}
\usepackage{amssymb}
\usepackage{pifont}
\newcommand{\cmark}{\ding{51}}
\newcommand{\xmark}{\ding{55}}
\usepackage{seqsplit}
\newcommand{\hash}[1]{{\ttfamily\seqsplit{#1}}}
\usepackage{tabularx}
\usepackage{comment}
\usepackage{lscape}

\definecolor{blackcolor}{rgb}{0,0,0}
\definecolor{codegray}{rgb}{0.5,0.5,0.5}
\definecolor{backcolour}{rgb}{0.95,0.95,0.92}
\usepackage{listings}

\lstdefinestyle{mystyle}{
    commentstyle=\color{codegray},
    keywordstyle=\color{blackcolor},
    numberstyle=\tiny\color{codegray},
    stringstyle=\color{codegray},
    basicstyle=\ttfamily\footnotesize,
    breakatwhitespace=false,         
    breaklines=true,                 
    captionpos=b,                    
    keepspaces=true,  
    frame = lines,
    numbers=left,                    
    numbersep=5pt,                  
    showspaces=false,                
    showstringspaces=false,
    showtabs=false,                  
    tabsize=2,
}
\usepackage{rotating} 
\usepackage{adjustbox}
\usepackage{multirow}

\usepackage[numbers]{natbib}
\usepackage{hyperref}
\begin{document}

\title{Dissecting contact tracing apps in the Android platform}
\date{}

\author[1]{Vasileios Kouliaridis}
\author[2]{Georgios Kambourakis}
\author[1]{Efstratios Chatzoglou}
\author[3]{Dimitrios Geneiatakis}
\author[4]{Hua Wang}

\affil[1]{Department of Information \& Communication Systems Engineering, University of the Aegean, Greece}
\affil[2]{European Commission, Joint Research Centre (JRC), 21027 Ispra (VA), Italy}
\affil[3]{European Commission, Directorate-General for Informatics, 1000 Bruxelles/Brussel, Belgium}
\affil[4]{Institute of Sustainable Industries and Liveable Cities, Victoria University, Melbourne, VIC 8001, Australia}

\maketitle


This work is published in PLOS ONE, DOI: https://doi.org/10.1371/journal.pone.0251867.
\\

\emph{Abstract}: Contact tracing has historically been used to retard the spread of infectious diseases, but if it is exercised by hand in large-scale, it is known to be a resource-intensive and quite deficient process. Nowadays, digital contact tracing has promptly emerged as an indispensable asset in the global fight against the coronavirus pandemic. The work at hand offers a meticulous study of all the official Android contact tracing apps deployed hitherto by European countries. Each app is closely scrutinized both statically and dynamically by means of dynamic instrumentation. Depending on the level of examination, static analysis results are grouped in two axes. The first encompasses permissions, API calls, and possible connections to external URLs, while the second concentrates on potential security weaknesses and vulnerabilities, including the use of trackers, in-depth manifest analysis, shared software analysis, and taint analysis. Dynamic analysis on the other hand collects data pertaining to Java classes and network traffic. The results demonstrate that while overall these apps are well-engineered, they are not free of weaknesses, vulnerabilities, and misconfigurations that may ultimately put the user security and privacy at risk.

\section{Introduction}
\label{S:Introduction}

Digital contact tracing mobile applications (apps) have lately emerged as an invaluable weapon in the hands of governments worldwide in the fight against the COVID-19 pandemic and future epidemics in general. Very quickly, a number of centralised or decentralised digital contact tracing architectures have been proposed \cite{Martin}, while even Google and Apple have aligned their forces to develop a joint privacy-preserving contact tracing framework API based on Bluetooth Low Energy (BLE), commonly known as Google-Apple Exposure Notification (GAEN) or Exposure Notifications System (ENS) \cite{GoogleApplecontacttracing}. For obvious reasons, the use of the Exposure Notification API is restricted to apps that have been developed by official public health agencies \cite{Restricted-API, XDA-2021}.

In the meanwhile, the topic of digital contact tracing has received considerable attention in the literature. In the following, we only refer to works that are pertinent to ours, intentionally leaving out others that are mostly concerned with frameworks \cite{DP-3T, PEPP-PT} and user privacy matters \cite{Fraunhofer, Avitabile}.

Hyunghoon et al. \cite{cho2020contact} focused on the Singapore's TraceTogether App and compared its characteristics against four alternative contact tracing systems, namely polling-based model, polling-based model with mixing, the use of a public database, and the exploitation of a private messaging system. They concentrated on the required computational infrastructure per system and on user privacy by considering a semi-honest model. Raskar et al. \cite{Raskar} outlined the different technological approaches to mobile-based contact tracing and elaborated on the risks that these technologies may pose to individuals and societies. Furthermore, the authors came up with a number of security enhancing approaches that can mitigate certain risks.

Nadeem et al. \cite{Nadeem} conducted a detailed overview of the existing contact tracing architectures as well as its privacy and security issues. Furthermore, the authors surveyed several contact tracing apps and analysed their privacy and security impact. Their work relies on public information about existing tracing apps, and the authors' own comprehension of the relevant protocols being developed for upcoming apps. Jinfeng \cite{Jinfeng} et al. presented a short survey of the various data regulations and technology protocols for contact tracing apps as well as a global overview of contact tracing app deployment. Martin et al. \cite{Martin} contributed a full-fledged comprehensive review of both the emerged contact tracing architectures and the contact tracing apps that have been already deployed by European countries. The discussion related to the apps is done in a rather high-level manner, i.e., concentrating on its installation and functionality, without touching on code analysis.

Recently, Samhi et al. \cite{samhi-2021} investigated COVID-related Android apps, in an attempt to empirically study their characteristics. Amongst others, the authors (a) generically reported on the top-10 most requested permissions by those apps, (b) examined the presence of privacy leaks by utilising the well-known, but rather obsolete (it has not been updated since 2016) \emph{FlowDroid-IccTA} tool, (c) checked whether each app is flagged by \emph{VirusTotal}, and (d) checked for potential misuse of crypto-APIs. They concluded that overall the apps do not seem to leak sensitive data. While this work is the one closer to ours, it is not devoted to privacy or security, thus the results reported by the authors remain mostly at a high-level and address a limited number of analysis axes.

Clearly, as digital tracing apps require mass acceptance in the population, and mass approval is largely based on perceived privacy, security and privacy design becomes a key concern for both policy makers and app architects in this field. Trang et al. \cite{Trang-2020} elaborate on this matter, and specifically concentrate on ways to attain mass acceptance of tracing apps depending on the context at hand. They identify three groups of citizens with different tendency for acceptance, namely critics, undecided, and advocates. Interestingly, they argue that only one set of tracing app specifications should be formulated for all citizens, and hence an exact targeting of diverse groups is not practical. Therefore, policy makers should first decide if the majority of the population can be assigned to one of the three groups or these groups are somewhat equally split. In the former case, their findings demonstrate that critics respond to societal-benefit appeals and privacy design, and thus policy makers must concentrate on communicating societal-benefit appeals and guarantee minimal exposure to privacy risks. Moreover, among others, although privacy design does play a significant role for undecided citizens, convenience in app usage seems to be more relevant to this group. For advocates, the authors' findings suggest that ``none of the benefit appeals is superior in achieving mass acceptance'', so ``policy makers may consider deprioritizing privacy and convenience as advocates would not penalize sacrifices to privacy and convenience''. However, if the population is roughly equally split among the three groups, policy makers should lean toward a carefully app specification set that fulfills the expectations of multiple groups at the same time. That is, they need to specify the app for meeting as closely as possible the needs of the critics and reluctant ones, if these groups prevail; in this respect, privacy, security, and convenience of use should be the primary impetus for persuading more people to participate.

\emph{Our contribution:} The work at hand aims at scrutinizing both the visible and under-the-hood functionality of the already deployed official Android contact tracing apps by European countries. The term ``official'' means apps that have been developed by official public health agencies. As seen from table \ref{T:apps}, up to now, a total of twenty-six apps are available. We specifically seek to deliver answers to a couple of essential matters with reference to the security and privacy aspects of this sort of apps. Namely, do these apps (a) curtail their functionality to the strict minimum?, and (b) remain free of known misconfigurations, weaknesses, vulnerabilities, and privacy leaks? As mentioned previously, and especially in reference to the work by \cite{Trang-2020}, privacy and security aspects are decisive factors when it comes to the mass acceptance of this kind of apps in the population, while this issue is also directly linked to the trade-offs that policy makers face in responding to the urgent challenges posed by the pandemic.

Naturally, app analysis in search of security and privacy concerns is a dynamic and constant process, and this holds especially true for the swiftly introduced contact tracing apps. Given that, until reaching a mature state, these apps are quite frequently updated, and thus the results presented here may not exactly catch up with the very latest version per app. Nevertheless, overall, we argue that the outcomes of this work more or less reflect the general propensity and can be used as a basis for further research in this field. The emphasis is on promoting best practices that facilitate the minimization of the attack surface, prevent data leakage, and foster security and privacy by design in general. We deliver hands-on results stemming from a full-spectrum static analysis, and in addition, we identify: (a) potentially privacy-invasive APIs, (b) possible connections to external URLs, i.e., not API-related URLs that the app might try to connect to, (c) configurations, weaknesses and vulnerabilities that may exist in the app's manifest file or code. Moreover, all apps have been dynamically analyzed by means of dynamic instrumentation. The outcomes in this case are not momentous, that is, no striking, unconventional, or covert behavior has been perceived, but mostly they serve as a verification step for the static analysis results. However, given that dynamic instrumentation basically depends on method hooking, and therefore needs intensive interaction with the running app, the aforementioned result can only be considered as rudimentary and more work is needed to examine and understand the exact behavior of the code. On top of that, in view of the fact that the great majority of such apps exploit GAEN \cite{GoogleApplecontacttracing}, we position each of them vis-\`a-vis a baseline privacy policy as defined by the above-mentioned framework.

Given the above, to the best of our knowledge, the current work is the first to elaborately profile the behavior of each of these apps after delving into its code both statically and dynamically. That is, in contrast to our work, all previous contributions in the literature deal with the security, privacy, or usability characteristics of the various hitherto deployed apps either in a high-level, incomplete manner or based on information made available in the app's website or elsewhere. 



The rest of this paper is structured as follows. The next section details on the methodology used to collect and analyze the apps. A first-tier static analysis focusing on apps' permissions and API calls is given in section \ref{S:High:Level:Static:Analysis}. A deeper static analysis targeting on a range of possible weaknesses and vulnerabilities is delivered in section \ref{S:Vulnerability:Analysis}. The penultimate section details on the results of dynamic analysis, while section \ref{S:Conclusions} concludes and gives pointers to future work.

\begin{table}[htpb]
\centering
\scriptsize
\begin{tabular}{|l|l|l|c|c|}
\toprule 
\textbf{App's name} & \textbf{Country} & \textbf{Analysed version} & \textbf{GAEN} & \textbf{Open-source}\\
\midrule

Stopp Corona \cite{O:Stopp:Corona} & Austria & 2.0.8.1133 & \cmark & \cmark \\
Coronalert \cite{O:Coronalert} & Belgium & 1.11.2 & \cmark & \cmark \\
ViruSafe \cite{O:Virusafe} & Bulgaria & 1.0.3 & \xmark\textsuperscript{$\ddagger$} & \cmark \\
Stop COVID-19 \cite{O:Stop:COVID:19} & Croatia & 2.2.0 & \cmark & \cmark \\
CovTracer \cite{Cyprus:New} & Cyprus & 1.3.2 & \cmark\textsuperscript{\textdagger} & \cmark \\
eRouska \cite{O:eRouska} & Czech Republic & 2.2.687 & \cmark\textsuperscript{*} & \cmark \\ 
Smittestop \cite{O:Smittestop} & Denmark & 2.1.1 & \cmark & \xmark \\
Hoia \cite{O:Hoia}& Estonia & 1.0.8 & \cmark & \xmark \\
Koronavilkku \cite{O:Koronavilkku} & Finland & 2.0.2 & \cmark & \cmark \\
TousAntiCovid \cite{O:TousAntiCovid} & France & 2.2.3 & \xmark\textsuperscript{$\ddagger$} & \xmark \\
Corona-Warn-App \cite{O:Corona:Warn:App}& Germany & 1.10.1 & \cmark & \cmark \\
VirusRadar \cite{O:VirusRadar}& Hungary & 1.0.0 & \xmark\textsuperscript{$\ddagger$} & \xmark \\
COVID Tracker \cite{O:COVID:Tracker} & Ireland & 1.0.4 & \cmark & \cmark \\
Immuni \cite{O:Immuni}& Italy & 2.2.1 & \cmark & \cmark \\
Apturi Covid \cite{O:Apturi:Covid}& Latvia & 1.1 & \cmark & \cmark \\
Korona Stop LT \cite{O:Korona:Stop:LT}& Lithuania & 1.1.1 & \cmark & \xmark \\
COVID Alert \cite{O:COVIDAlert}& Malta & 1.3.5 & \cmark & \cmark \\
CoronaMelder \cite{O:CoronaMelder}& Netherlands & 1.2.2 & \cmark & \cmark \\
Smittestopp \cite{O:Smittestopp}& Norway & 1.0.3 & \cmark & \cmark \\
STOP COVID - ProteGO Safe \cite{O:STOP:COVID:ProteGO}& Poland & 4.9.0 & \cmark  & \cmark \\
STAYAWAY COVID \cite{O:STAYAWAY:COVID}& Portugal & 1.1.2 & \cmark & \cmark \\
ZostanZdravy \cite{O:ZostanZdravy} & Slovakia & 1.1.0 & \xmark\textsuperscript{$\ddagger$\#} & \cmark \\
\#OstaniZdrav \cite{O:OstaniZdrav}& Slovenia & 1.10.1 & \cmark & \xmark \\
Radar COVID \cite{O:Radar:COVID}& Spain & 1.2.0 & \cmark & \cmark \\
SwissCovid \cite{O:Swiss:Covid}& Switzerland & 1.3.1 & \cmark & \cmark \\
NHS COVID-19 \cite{O:NHS:COVID:19}& UK & 4.3 & \cmark & \cmark \\

\bottomrule
\end{tabular}
\caption{Outline of the examined apps. The interested reader can also refer to \cite{XDA-2021}, which gathers all GAEN-based apps worldwide. \textsuperscript{$\ddagger$}Centralised/proprietary approach. \textsuperscript{\textdagger}Formerly known as CovTracer. It embraced GAEN on Dec. 2020.
\textsuperscript{*}From ver. 2.0 onward. Prior versions were based on a centralised framework. \textsuperscript{\#}The official website at https://korona.gov.sk/en/ provides a dead link to Google Play Store.}
\label{T:apps}
\end{table}

\section{Methodology}
\label{S:methodology}

Twenty six official apps were collected from the Google Play Store \cite{Playstore} with freeze date as of February 2, 2021. All the essential pieces of information per app, including the analyzed version, whether the app is based on GAEN, and if its source code is publicly available are summarized in table \ref{T:apps}. The apps are sorted alphabetically by their country of origin, and the same sorting order is followed for the rest of the tables and figures across all the sections of this work. For a more detailed high-level presentation of each app, the interested reader may refer to the work of Martin et al. \cite{Martin} and \cite{O:EU:All:Apps}.

\begin{figure}[ht]
 \begin{center}
        \includegraphics[width=1\linewidth]{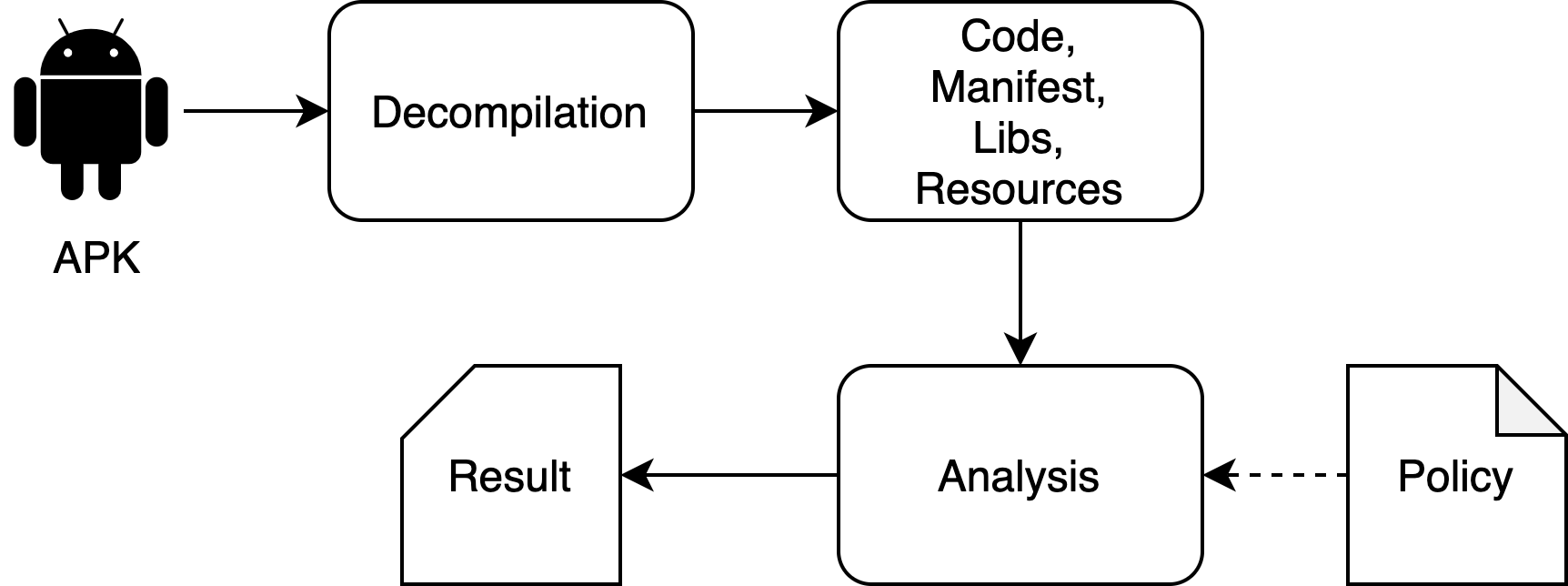}
        \caption{Overview of the followed methodology.}
 \label{F:Analysis:Procedure}
 \end{center}
\end{figure}

As shown in figure \ref{F:Analysis:Procedure}, two axes of analysis were followed. The first, scrutinizes each app statically, while the second examines the app by running it. Specifically, the static axis incorporates a number of analysis steps, including (a) sensitive permissions and API calls, and third-party trackers, which target mainly the privacy of the end-user, and (b) misconfigurations, weaknesses and vulnerabilities, which focus on the security of the app. For instance, the latter step inspects the code of each app for possibly identifying Common Weakness Enumerations (CWEs), Common Vulnerabilities and Exposures (CVEs), misconfigurations attributed to the use of shared libraries, etc. This step also includes a basic taint analysis.

Static analysis employed three different tools. Two of them, namely \emph{Androtomist} \cite{Androtomist} and \emph{MobSF} \cite{MobSF} are open-source, while the other, namely \emph{Ostorlab} \cite{O:Ostorlab}, utilized only for outdated software component analysis and taint analysis, is a software-as-a-service (SaaS) product. Specifics on these tools are given in the respective sections. When looking for weaknesses and vulnerabilities, we also relied on the methodology set out in the OWASP mobile security testing guide \cite{O:OWASP:guide}.

On the other hand, the dynamic analysis axis concentrates on the use of Java classes and network traffic. As already mentioned, this axis of analysis has been carried out by means of dynamic instrumentation, i.e., the instrumentation, namely, method hooking, is added during the app execution in a just-in-time (JIT) approach. For this task, the Androtomist tool was exploited along with the Nvidia Shield K tablet \cite{NvidiaShield}, running on Android 7.0. Note that this device is SIM-less, which, depending on the case, may affect the proper running of apps that require cellular services. To cover all possible functionality, each app was exercised by hand, that is, not via the use of a UI/app exerciser.

\section{High-level static analysis}
\label{S:High:Level:Static:Analysis}

Each app has been statically analyzed following the step-by-step procedure illustrated in Figure~\ref{F:Analysis:Procedure}. Namely, the corresponding android package (APK) archive has been extracted and then decompiled using the \emph{apktool} \cite{Apktool}. Next, (a) the manifest file has been unpacked and compared against a baseline policy given in subsection \ref{SS:Baseline:Policy}, and (b) the API calls have been retrieved from the \emph{smali} files. The results have been cross-checked via the Androtomist tool \cite{Androtomist}. At minimum, this kind of analysis offers insight on whether the app:

\begin{enumerate}
    \item Follows the least privilege principle regarding its declared permissions.
    \item Might enable privacy-sensitive, that is, ``dangerous'' operations as designated by the Android API \cite{API}.
    \item Contains API calls that do not coincide with the requested permissions.
\end{enumerate}

Furthermore, we provide an overall evaluation of the security level of the external network domains each app might connect to, that is, every URL found in its code. This assessment was done with the aid of the ``SSL Server Test'' as offered by Qualys SSL labs \cite{qualisssllabs}. Precisely, as detailed in \cite{qualisssllabstestguide}, this test thoroughly examines the configuration of any TLS web server based on four main criteria, namely, certificate, protocol support, key exchange, and cipher strength. The results vary from ``A+'' (exceptional configuration) to ``F'' (vulnerabilities exist). Also, the letter ``T'' means that the website certificate is not trusted.

\subsection{Baseline policy}
\label{SS:Baseline:Policy}

Assuming the use of GAEN \cite{GoogleApplecontacttracing}, as a baseline, contact tracing apps based on BLE should at the bare minimum ask for the following two ``normal'' permissions \cite{policy, D3PTManifest, ENS-2}:

\begin{itemize}

\item BLUETOOTH: This permission is necessary for enabling any Bluetooth communication, i.e., either requesting/accepting a connection or transferring data.

\item INTERNET: It enables the app to perform network operations. Note that this permission goes hand in hand with the ACCESS\_NETWORK\_STATE one.

\end{itemize}

According to \cite{ENS-2}, a GAEN-based contact tracing app does not need and cannot include either BLUETOOTH\_ADMIN, ACCESS\_FINE\_LOCATION, or ACCESS\_COARSE\_LOCATION. The latter two permissions are classified as ``dangerous'' in the Android API. Recall that, in contrast to a ``normal'' permission, every higher-risk one necessitates prompting the user.

\begin{itemize}

\item BLUETOOTH\_ADMIN. It is required to trigger device discovery or manipulate the Bluetooth settings. Note that if an app requires this permission, then the app must also declare the BLUETOOTH one.

\item ACCESS\_FINE\_LOCATION. This permission allows the app to access the precise location of the device via the use of GPS, Wi-Fi, or mobile cell data. On devices running Android 6.0 to 10, the ENS relies on Bluetooth scanning, which in turn mandates the device \emph{location setting} to be turned on \cite{locationExplained}. It should be pointed out that according to \cite{policy}, the smartphone's location setting should be enabled: ``The system uses this to scan for Bluetooth signals. The system does not collect or track your location''. And in \cite{policy}, public health authority ``apps that use the ENS are not permitted to request permission to use your device location''. On the other hand, for devices running Android 11, ENS does not mandate the smartphone's location setting to be enabled: ``We're making this update for ENS only, given that ENS has been designed in such a way that neither the system nor the apps using it can infer device location through Bluetooth scanning, and apps that are allowed to use ENS are subject to additional policies that disallow automatic collection of location'' \cite{policy3}.

\item ACCESS\_COARSE\_LOCATION is potentially privacy-invasive as it allows an app to access the approximate location of the device through the use of either or both Wi-Fi and mobile cell data. The accuracy provided to the app is approximately equivalent to a city block.

\end{itemize}


As already pointed out, API calls per app have been also tracked down. However, in this case, instead of providing a baseline, which is considered impractical, this section only identifies certain API calls that may pose a threat to user privacy. On top of that, this enables one to detect API calls that are in conflict with the requested permissions. Precisely, the focus is on the following list of API calls:

\begin{itemize}

    \item \emph{\seqsplit{android/telephony/TelephonyManager;$\,\to\,$getNetworkOperator()}} returns the mobile country code (MCC) and mobile network code (MNC) of the current registered operator.
    \item \emph{\seqsplit{android/telephony/TelephonyManager;$\,\to\,$getNetworkOperatorName()}} returns the alphabetic name of the current registered operator.
    \item \emph{\seqsplit{android/telephony/TelephonyManager;$\,\to\,$getLine1Number()}} returns the phone number string for line 1, say, the Mobile Station International Subscriber Directory Number (MSISDN) for a GSM phone for a specific subscription. It requires at least one of the following permissions READ\_PHONE\_STATE, READ\_SMS, or READ\_PHONE\_NUMBERS.
    \item \emph{\seqsplit{android/telephony/TelephonyManager;$\,\to\,$getSimCountryIso()}} returns the SIM provider's country code.
    \item \emph{\seqsplit{android/telephony/TelephonyManager;$\,\to\,$getSimOperatorName()}} returns the service provider name.
    \item \emph{\seqsplit{android/telephony/TelephonyManager;$\,\to\,$getCellLocation()}} returns the current location of the device and requires the ACCESS\_FINE\_LOCATION permission. This method was deprecated in Android v8.
    \item \emph{\seqsplit{android/location/LocationManager;$\,\to\,$getLastKnownLocation()}} gets the last known (cached in memory) location, if any, from the given provider. It requires either the ACCESS\_COARSE\_LOCATION or ACCESS\_FINE\_LOCATION permission.
    \item \emph{\seqsplit{android/location/LocationManager;$\,\to\,$requestLocationUpdates()}} is used to register for location updates from the given provider. It presupposes the same permission as the getLastKnownLocation() one.
    \item \emph{\seqsplit{android/location/Location;$\,\to\,$getLatitude()}} and \emph{\seqsplit{android/location/Location;$\,\to\,$getLongitude()}} API calls are used to obtain the latitude and longitude of the device, respectively.
    \item \emph{\seqsplit{android/hardware/Camera;$\,\to\,$open()}} is used to access a particular hardware camera. It presupposes the CAMERA permission.
    \item \emph{\seqsplit{android/hardware/camera2/CameraManager}} is a system service manager for detecting, characterising, and connecting to CameraDevice.

\end{itemize}

\subsection{Analysis of apps}
\label{SS:Analysis:of:apps}

This section outlines all noteworthy results per app in regard to this first level of static analysis. Extracting the security permissions listed in the \emph{AndroidManifest.xml} file of an app is a principal step towards understanding its general behavior. Moreover, the lookup for potentially privacy-invasive API calls in the app's code can on the one hand offer supplementary information about higher-risk actions the app may perform, and on the other, reveal whether the identified calls coincide with the requested permissions. Table \ref{T:permissions:baseline} summarizes the requested permissions per app with reference to the baseline policy. Additionally, the fourth column of the table shows any extra ``dangerous'' permission identified in each analyzed app.

\begin{itemize}

\item \emph{ViruSafe} does not rely on BLE for contact tracing, but on GPS. Naturally, for this app, the baseline policy given in section \ref{SS:Baseline:Policy} does not apply. Also, \emph{TousAntiCovid}, \emph{VirusRadar}, and \emph{ZostanZdravy} use BLE in conjunction with GPS.

\item With reference to the data minimization principle, all GAEN-based apps seem to abide by the baseline policy. This ensures that only the data which are strictly essential for the running of the service are gathered. Interestingly, only two apps declare an additional dangerous permission, namely CAMERA, on top of those mandated by the baseline policy.

\item Regarding ``dangerous'' permissions for not GAEN-based apps, ACCESS\_COARSE\_LOCATION, ACCESS\_FINE\_LOCATION, CAMERA, and ACCESS\_BACKGROUND\_LOCATION are exploited by 2, 4, 1, and 2 of them, respectively. Note that the latter permission allows an app to access location in the background, and it goes hand in hand with either ACCESS\_COARSE\_LOCATION or ACCESS\_FINE\_LOCATION.

\item As observed from table \ref{T:external:URLs}, the code of all the apps but 5 contain at least one external, not API-related, URL where the app might try to connect to. The security analysis of the domains of these hyperlinks designate that most of them are classified in the range from A+ to B according to \cite{qualisssllabs}. Specifically, from the total of 49 URLs, about 69\% were classified as A+ or A, which is definitely on the plus side. On the other hand, an approximately 20\% were ranked as B, three URLs have received a ``T'' score, and another one an ``F''. No less important, 11 apps or $\approx$41\% contain only A+ of A-rated URLs.

\begin{table}[htpb]
\footnotesize
\begin{tabular}{ |p{4cm}|c|c|p{4cm}| }
\hline
\textbf{App} & \textbf{BLUETOOTH} & \textbf{INTERNET} & \textbf{Extra permissions} \\
\hline
Stopp Corona        & +  & + & -- \\
Coronalert          & +  & + & -- \\
ViruSafe            & -- & + & \hash{ACCESS\_BACKGROUND\_LOCATION, ACCESS\_FINE\_LOCATION} \\
Stop COVID-19       & +  & + & -- \\
CovTracer           & +  & + & -- \\
eRouska             & +  & + & -- \\
Smittestop          & +  & + & -- \\
Hoia                & +  & + & -- \\
KoronaVilkku        & +  & + & -- \\
TousAntiCovid       & +  & + & \hash{CAMERA, ACCESS\_COARSE\_LOCATION, ACCESS\_FINE\_LOCATION}  \\
Corona-Warn-App     & +  & + & \hash{CAMERA} \\
VirusRadar          & +  & + & \hash{ACCESS\_COARSE\_LOCATION, ACCESS\_FINE\_LOCATION} \\
COVID Tracker       & +  & + & -- \\
Immuni              & +  & + & -- \\
Apturi Covid        & +  & + & -- \\
Korona Stop LT      & +  & + & -- \\
COVID Alert          & +  & + & -- \\
CoronaMelder        & +  & + & -- \\
Smittestopp         & +  & + & -- \\
STOP COVID - ProteGO Safe & +  & + & -- \\
STAYAWAY COVID      & +  & + & -- \\
ZostanZdravy        & +  & + & \hash{ACCESS\_FINE\_LOCATION, ACCESS\_BACKGROUND\_LOCATION}  \\
\#OstaniZdrav       & +  & + & -- \\
Radar COVID         & +  & + & -- \\
SwissCovid          & +  & + & -- \\
NHS COVID-19        & +  & + & \hash{CAMERA} \\
\hline
Total               & 25 & 26 & -- \\
\hline
\end{tabular}
\caption{Overview of requested permissions per app vis-\`a-vis the baseline.} 
\label{T:permissions:baseline}
\end{table}

\begin{table}[htpb]
\centering
\begin{adjustbox}{width=1\textwidth}
\begin{tabular}{|l|l|l|l|}
\hline
\multicolumn{1}{|l|}{\textbf{App}} & \multicolumn{1}{|l|}{\textbf{Number of URLs}} & \multicolumn{1}{|l|}{\textbf{Domain ratings}} & \multicolumn{1}{|l|}{\textbf{Domains}} \\ \hline
\multirow{3}{*}{Stopp Corona}              & \multirow{3}{*}{3} & A   & https://cdn.prod-rca-coronaapp-fd.net  \\
                                           &                    & A   & https://app.prod-rca-coronaapp-fd.net  \\
                                           &                    & A+  & https://sms.prod-rca-coronaapp-fd.net  \\\hline
\multirow{5}{*}{Coronalert}                & \multirow{5}{*}{5} & A   & https://c19statcdn-prd.ixor.be         \\
                                           &                    & A   & coronalert-prd.ixor.be                 \\
                                           &                    & A   & c19statcdn-prd.ixor.be                 \\
                                           &                    & A   & c19-submission-prd.ixor.be             \\
                                           &                    & A   & c19-verification-prd.ixor.be           \\\hline
ViruSafe                                   & 1                  & A   & https://virusafe.io/                   \\\hline
Stop COVID-19                              & 0                  & --  & --                                     \\\hline
\multirow{2}{*}{CovTracer}                 & \multirow{2}{*}{2} & A   & https://kiosapps.ucy.ac.cy             \\
                                           &                    & B   & covtracer.dmrid.gov.cy                 \\\hline
eRouska                                    & 1                  & A+  & https://erouska.cz                     \\\hline
Smittestop                                 & 0                  & --  & --                                     \\\hline
\multirow{4}{*}{Hoia}                      & \multirow{4}{*}{4} & A   & https://www.digilugu.ee                \\
                                           &                    & A+  & https://www.terviseamet.ee             \\
                                           &                    & A+  & https://hoia.me                        \\
                                           &                    & A   & https://enapi.sm.ee                    \\\hline
\multirow{2}{*}{KoronaVilkku}              & \multirow{2}{*}{2} & A   & https://omaolo.fi                      \\
                                           &                    & A+  & https://repo.thl.fi                    \\\hline
\multirow{5}{*}{TousAntiCovid}             & \multirow{5}{*}{5} & B   & https://api.stopcovid.gouv.fr          \\
                                           &                    & B   & https://tacw.tousanticovid.gouv.fr     \\
                                           &                    & T   & https://stopcovid.gouv.fr              \\
                                           &                    & T   & https://app.stopcovid.gouv.fr          \\
                                           &                    & A   & https://tac.gouv.fr                    \\\hline
Corona-Warn-App                            & 1                  & N/A (dead link) & https://submission.coronawarn.ap       \\\hline
\multirow{2}{*}{VirusRadar}                & \multirow{2}{*}{2} & B   & https://f-droid.org                    \\
                                           &                    & A+  & https://github.com                     \\\hline
COVID Tracker                              & 0                  & --  & --                                     \\\hline
Immuni                                     & 1                  & A+  & https://www.immuni.italia.it           \\\hline
\multirow{3}{*}{Apturi Covid}              & \multirow{3}{*}{3} & A+  & https://apturicovid-files.spkc.gov.lv  \\
                                           &                    & A+  & apturicovid-api.spkc.gov.lv            \\
                                           &                    & A+  & apturicovid-files.spkc.gov.lv          \\\hline
\multirow{3}{*}{Korona Stop LT}            & \multirow{3}{*}{3} & B   & https://download.koronastop.lt         \\
                                           &                    & B   & https://submission.koronastop.lt       \\
                                           &                    & B   & https://verification.koronastop.lt     \\\hline
\multirow{4}{*}{COVID Alert}                & \multirow{4}{*}{4} & B   & https://tools.ietf.org                 \\
                                           &                    & A   & https://data.ws.covidalert.gov.mt      \\
                                           &                    & A   & https://authz.ws.covidalert.gov.mt     \\
                                           &                    & A   & https://config.ws.covidalert.gov.mt    \\\hline
\multirow{2}{*}{CoronaMelder}              & \multirow{2}{*}{2} & A+  & https://coronamelder-api.nl            \\
                                           &                    & B   & https://productie.coronamelder-dist.nl \\\hline
Smittestopp                                & 1                  & A   & https://in.appcenter.ms                \\\hline
\multirow{2}{*}{STOP COVID - ProteGO Safe} & \multirow{2}{*}{2} & A+  & https://safesafe.app                   \\
                                           &                    & A+  & https://v4.safesafe.app                \\\hline
\multirow{2}{*}{STAYAWAY COVID}            & \multirow{2}{*}{2} & F   & https://stayaway.min-saude.pt          \\
                                           &                    & A   & https://stayaway.incm.pt               \\\hline
ZostanZdravy                               & 1                  & B    & https://t.mojeezdravie.sk              \\\hline
\#OstaniZdrav                              & 0                  & --  & --                                     \\\hline
Radar COVID                                & 1                  & A    & https://radarcovid.covid19.gob.es      \\\hline
\multirow{3}{*}{SwissCovid}                & \multirow{3}{*}{3} & A    & https://www.pt1.bfs.admin.ch           \\
                                           &                    & T    & https://www.pt.bfs.admin.ch            \\
                                           &                    &  A   & https://codegen-service.bag.admin.ch   \\\hline
NHS COVID-19                               & 0                  & --  & --          \\ \hline
Total                                   & 49 & 13xA+, 21xA, 10xB, 3xT, 1xF & \\
\hline
\end{tabular}
\end{adjustbox}
\caption{TLS server security level rating of external URLs per app.}
\label{T:external:URLs} 
\end{table}

\end{itemize}

As already pointed out, API calls per app have been also extracted with the aim of identifying certain call to methods which may pose a threat to the user's privacy. Specifically, the API calls listed in section \ref{SS:Baseline:Policy} have been grouped in table \ref{T:ApiCall-groups} into three categories, namely cellular network, location, and camera. It is noteworthy that API calls related to the cellular network may also expose the user's location. For example, the phone number (getLine1Number()) or SIM operator name (getSimOperatorName()) may reveal the user's country, while getCellLocation() returns the current location of the device. Precisely, as outlined in table \ref{T:apicalls:baseline}, the following key observations have been made.

\begin{itemize}

\item All apps include \emph{\seqsplit{android/location/LocationManager;$\,\to\,$getLastKnownLocation}}, which gets the last known location, if any, from the given provider. On top of that, all apps include the
\emph{\seqsplit{android/location/Location;$\,\to\,$getLatitude}} and
\emph{\seqsplit{android/location/Location;$\,\to\,$getLongitude}} API calls, which are used to get the latitude and longitude of the device. Nevertheless, as explained in \ref{SS:Baseline:Policy} for the ACCESS\_FINE\_LOCATION, according to \cite{locationExplained}, GAEN-based apps are not permitted to request permission to use the device location. Also, according to the Android API, only apps which declare the ACCESS\_COARSE\_LOCATION or ACCESS\_FINE\_LOCATION permissions are able to get device location data. So, overall, the aforementioned three calls can be only carried out by apps which do not follow GAEN.

\item \emph{Corona-Warn-App-App}, \emph{TousAntiCovid}, \emph{NHS COVID-19}, \emph{Apturi Covid}, \emph{Coronalert}, and \emph{\#OstaniZdrav} include \emph{\seqsplit{android/hardware/Camera;$\,\to\,$open}}, which creates a new Camera object to access the first back-facing camera on the device, if any. For the three first of the aforementioned apps this API call coincides with the requested permissions. Nevertheless, for the rest of them, this call is unjustifiable vis-\`a-vis the requested permissions, and anyway cannot be executed. Interestingly, according to an article from \cite{bbc}, Apple and Google recently refused to publish an updated version of the \emph{NHS COVID-19} app. The updated version was asking users to upload logs of venue check-ins, via barcode scans, if they tested positive for the virus. The reason behind this extra functionality would be to warn others. To enable this operation, an app would require permission for the camera, as well as the aforementioned API call. Both conditions are met for the three first of the above-mentioned apps.

\item
\emph{CovTracer}, \emph{COVID Tracker}, and \emph{Smittestopp} include the \emph{\seqsplit{android/telephony/TelephonyManager;$\,\to\,$getNetworkOperatorName}} call, which returns the alphabetical name of the current registered operator. As already pointed out, this may reveal the user's country. According to the API, this method does not require any special permission, thus all these apps can indeed get the relevant information.

\end{itemize}

\begin{table}[htpb]
\small
\centering
\begin{tabular}{|l|l|}
\hline
\textbf{Relevant System} & \textbf{API Calls}  \\ \hline
\multirow{6}{*}{Cellular Network} & android/telephony/TelephonyManager;$\,\to\,$getNetworkOperatorName() \\
 & android/telephony/TelephonyManager;$\,\to\,$getNetworkOperator() \\
 & android/telephony/TelephonyManager;$\,\to\,$getLine1Number() \\
 & android/telephony/TelephonyManager;$\,\to\,$getSimOperatorName() \\
 & android/telephony/TelephonyManager;$\,\to\,$getSimCountryIso() \\
 & android/telephony/TelephonyManager;$\,\to\,$getCellLocation() \\ \hline
\multirow{4}{*}{Location} & android/location/LocationManager;$\,\to\,$getLastKnownLocation() \\
 & android/location/LocationManager;$\,\to\,$requestLocationUpdates() \\
 & android/location/Location;$\,\to\,$getLatitude() \\
 & android/location/Location;$\,\to\,$getLongitude() \\ \hline
Camera  & android/hardware/Camera;$\,\to\,$open() \\ & android/hardware/camera2/CameraManager;$\,\to\,$* \\ \hline
\end{tabular}
\caption{Categorisation of sensitive API calls discovered in the apps. *All class methods.}

\label{T:ApiCall-groups}
\end{table}

\begin{table}[htpb]
\centering
\footnotesize

\begin{tabular}{ |p{4.6cm}|c|c|c|c| }
\hline
\textbf{App} & \textbf{Cellular Network} & \textbf{Location} & \textbf{Camera} \\
\hline
Stopp Corona        & -- & +* & -- \\
Coronalert          & -- & +* & +* \\
ViruSafe            & -- & + & -- \\
Stop COVID-19       & -- & +* & -- \\
CovTracer           & +  & +* & -- \\
eRouska             & -- & +* & -- \\
Smittestop          & --  & +* & -- \\
Hoia                & -- & +* & -- \\
KoronaVilkku        & -- & +* & -- \\
TousAntiCovid       & -- & + & + \\
Corona-Warn-App     & -- & +* & + \\
VirusRadar          & -- & + & -- \\
COVID Tracker       & + & +* & -- \\
Immuni              & -- & +* & -- \\
Apturi Covid        & -- & +* & +* \\
Korona Stop LT      & -- & +* & -- \\
COVID Alert          & -- & +* & -- \\
CoronaMelder        & -- & +* & -- \\
Smittestopp         & + & +* & -- \\
STOP COVID - ProteGO Safe & -- & +* & -- \\
STAYAWAY COVID      & -- & +* & -- \\
ZostanZdravy        & -- & + & -- \\
\#OstaniZdrav       & -- & +* & +* \\
Radar COVID         & -- & +* & -- \\
SwissCovid          & -- & +* & -- \\
NHS COVID-19        & -- & +* & + \\
\hline
Total               & 3 & 26 & 6 \\
\hline
\end{tabular}
\caption{Potentially privacy-invasive API call groups per app. The asterisk indicates that at least one API call does not coincide with the permissions requested by this app.} 
\label{T:apicalls:baseline}
\end{table}






\section{Low-level static analysis}
\label{S:Vulnerability:Analysis}

To investigate further each examined app, and for subsections \ref{SS:SignerCert:APKiD:Network security} to \ref{SS:Shared:Library:Binary:Analysis}, we relied on the well-known Mobile Security Framework (MobSF) in v3.2.4 \cite{MobSF}. MobSF is capable of performing both static and dynamic analysis and is used as a pen-testing, malware analysis, and security assessment framework. It is also one of the all-in-one tools recommended by the OWASP Mobile Security Testing Guide \cite{O:OWASP:guide}.

As seen in table \ref{T:Vulnerabilities:per:app}, the focus of this part of study is on signer certificate information, APKiD, network security, code analysis aiming at divulging CWEs, tracker analysis, manifest analysis, and shared library binary analysis. For all these categories, and for the sake of conciseness, we only consider high severity (or high value according to the common weakness scoring system) weaknesses. For extracting the above-mentioned pieces of information, MobSF decompiles the provided APK using Dex to Java decompiler \emph{(Jadx)} \cite{O:JADX}; code de-obfuscation processes are applicable to this step as well.

\begin{sidewaystable}
\centering
\begin{adjustbox}{width=1\textwidth}
\begin{tabular}{|l|l|l|l|l|l|l|l|l|l|l|l|l|l|l|l|}
\hline
\textbf{App Name} &
  \textbf{Signer certificate information} &
  \textbf{Janus} &
  \textbf{Network Security} &
  \textbf{CWE-330} &
  \textbf{CWE-276} &
  \textbf{CWE-532} &
  \textbf{CWE-312} &
  \textbf{CWE-89} &
  \textbf{CWE-327} &
  \textbf{CWE-295} &
  \textbf{CWE-749} &
  \textbf{CWE-919}\\ \hline
Stopp Corona &
  -- & 
  + & 
  -- & 
  + & 
  + & 
  -- & 
  -- & 
  -- & 
  -- & 
  -- & 
  -- & 
  -- \\ 
Coronalert &
  -- & 
  + & 
  --
   & 
  + & 
  + & 
  + & 
  + & 
  + & 
  -- & 
  + & 
  -- & 
  -- \\ 
ViruSafe &
  -- & 
  + & 
  -- & 
  + & 
  -- & 
  + & 
  + & 
  + & 
  AES ECB & 
  -- & 
  -- & 
  -- \\ 
Stop COVID-19 &
  -- & 
  + & 
  -- & 
  + & 
  + & 
  + & 
  -- & 
  -- & 
  -- & 
  -- & 
   --& 
  -- \\ 
CovTracer &
  -- & 
  + & 
  -- & 
 +  & 
  -- & 
 +  & 
 +  & 
 +  & 
  MD5, AES-ECB & 
  -- & 
 --  & 
 --  \\ 
eRouska &
 --  & 
  + & 
 --  & 
  + & 
  -- & 
  + & 
  + & 
   & 
  -- & 
  -- & 
  --& 
  -- \\ 
Smittestop &
  -- & 
  + & 
  -- & 
  -- & 
  + & 
  + & 
  -- & 
  -- & 
  -- & 
  -- & 
  -- & 
  -- \\ 
Hoia &
  -- & 
  + & 
  -- & 
  + & 
  -- & 
  + & 
  + & 
  + & 
  -- & 
  -- & 
  -- & 
  -- \\ 
Koronavilkku &
  -- & 
  + & 
  -- & 
  + & 
  + & 
  + & 
  + & 
  + & 
  -- & 
  -- & 
  -- & 
  -- \\ 
TousAntiCovid &
  -- & 
  + & 
  -- & 
  + & 
  + & 
  + & 
  + & 
  -- & 
  -- & 
  -- & 
  -- & 
  -- \\ 
Corona-Warn-App &
  -- & 
  + & 
  -- & 
  + & 
  + & 
  + & 
  -- & 
  + & 
  MD5, SHA1 & 
  + & 
  -- & 
   --\\ 
VirusRadar &
  -- & 
  + & 
  -- & 
  + & 
  + & 
  + & 
  + & 
  -- & 
  -- & 
  --& 
  -- & 
   --\\ 
COVID Tracker &
  -- & 
  + & 
  -- & 
  + & 
  ++ & 
  + & 
  + & 
  + & 
  SHA1 & 
  -- & 
  -- & 
  -- \\ 
Immuni &
  -- & 
  + & 
  -- & 
  + & 
  + & 
  + & 
  + & 
  -- & 
  -- & 
  + & 
  -- & 
  -- \\ 
Apturi Covid &
  -- & 
  + & 
  -- & 
  + & 
  + & 
  + & 
  + & 
  -- & 
  -- & 
  -- & 
  -- & 
  -- \\ 
Korona Stop LT &
  -- & 
  + & 
  -- & 
  + & 
  -- & 
  + & 
  -- & 
  + & 
  SHA1 & 
  + & 
  -- & 
  -- \\ 
COVID Alert &
  SHA1withRSA (SHA256withRSA) & 
  + & 
  -- & 
  + & 
  -- & 
  -- & 
  + & 
  + & 
  -- & 
  -- & 
  -- & 
  -- \\ 
CoronaMelder &
  -- & 
  + & 
  -- & 
  + & 
  ++ & 
  + & 
  + & 
  + & 
  -- & 
  -- & 
  -- & 
   --\\ 
Smittestopp &
  -- & 
  + & 
  -- & 
  + & 
  + & 
  + & 
  + & 
  + & 
  -- & 
  -- & 
   --& 
   --\\ 
STOP COVID - ProteGO Safe &
  -- & 
  + & 
  -- & 
  + & 
  -- & 
  + & 
  -- & 
  -- & 
  -- & 
  -- & 
  + & 
  -- \\ 
STAYAWAY COVID &
  -- & 
  + & 
  -- & 
  + & 
  -- & 
  + & 
  + & 
  + & 
  -- & 
  -- & 
  -- & 
  -- \\ 
ZostanZdravy &
  -- & 
  + & 
  \begin{tabular}[c]{@{}l@{}}Domain is insecurely configured \\and permits clear text traffic\end{tabular} & 
  + & 
  ++ & 
  + & 
  -- & 
 + & 
 MD5, SHA1, AES-ECB & 
  -- & 
  -- & 
  -- \\ 
\#OstaniZdrav &
  -- & 
  + & 
  -- & 
  + & 
  + & 
  + & 
  -- & 
  + & 
  MD5, SHA1 & 
  + & 
  -- & 
  -- \\ 
Radar COVID &
  -- & 
  + & 
  -- & 
  + & 
  -- & 
  + & 
  + & 
  + & 
  -- & 
  -- & 
  -- & 
  -- \\ 
SwissCovid &
  -- & 
  + & 
  -- & 
  + & 
  -- & 
  + & 
  + & 
  + & 
  -- & 
  -- & 
  -- & 
  -- \\ 
NHS Covid-19 &
  -- & 
  + & 
  \begin{tabular}[c]{@{}l@{}}Domain is insecurely configured \\and permits clear text traffic\end{tabular} & 
  + & 
  + & 
  + & 
  + & 
  -- & 
  -- & 
  -- & 
  -- & 
   +\\ 
\hline
Total & 1 & 26 & 2 & 25 & 16 & 24 & 18 & 17 & 7 & 5 & 1 & 1 \\
\hline
\end{tabular}
\end{adjustbox}
\caption{Potential weaknesses and other security issues per app.}
\label{T:Vulnerabilities:per:app}
\end{sidewaystable}

Moreover, the last two subsections of the current section are devoted to the use of outdated third-party software by the apps and to taint analysis, respectively. For both these undertakings, the Ostorlab tool was utilised. Ostorlab is a well-known software-as-a-service (SaaS) product to review the security and privacy of mobile apps. Note that for the sake of the reproducibility of the results, we employed the free-to-use ``community'' edition of the tool. To our knowledge, the same tool has been exploited in the context of similar researches \cite{A:Ostorlab:1:Mabo, A:Ostorlab:2:Imai, Chatzoglou-2021}.

\subsection{Signer Cert., APKiD, Network security}
\label{SS:SignerCert:APKiD:Network security}

Each APK is signed by the developer using a specific cryptographic hash function, say, SHA-1, and APK signature scheme version, say, v3. As observed from the second column of table \ref{T:Vulnerabilities:per:app}, one app indicates a diverse hash algorithm (SHA256) than the one actually used (SHA-1) to sign the app. Precisely, the algorithm in parenthesis indicates the hash algorithm declared in the app's manifest file, while the actual algorithm used is shown at the left.
Note that NIST deprecated the use of SHA-1 and suppressed its use for digital signatures in 2011 and 2013, respectively \cite{O:NIST:SHA1:MD5}. Precisely, if the app has been signed with the use of SHA-1, collisions may be possible. This means that any app signed with the corresponding algorithm is prone to attacks, including hijacking the app with phony updates or granting permissions to a malicious app. For example, the attacker can repackage the app after including malicious code in it. Then, if the signature validates, they could lure users to install the repacked app instead of the legitimate one.

Another vulnerability which is rooted to improper signature usage is commonly known as ``Janus'' (CVE-2017-13156) \cite{Janus:CVE}. It can be exploited if the v1 signature scheme (JAR signing) is used along with Android v5.0 (API 21) to v7.0 (API 25). Specifically, Janus leverages the possibility of adding extra bytes to APK and DEX files, without affecting the signature. As perceived from table \ref{T:Vulnerabilities:per:app}, all the examined apps are vulnerable to Janus, namely, they were signed under scheme v1 and support Android v6.
    
As presented in table \ref{T:Vulnerabilities:per:app}, network security analysis pinpointed one high severity, self-explanatory vulnerability, namely, the ``domain is insecurely configured and permits clear text traffic''. It is important to note that this issue refers to the network security configuration xml file an app may designate via a specific entry \textless application android:networkSecurityConfig=``''\textgreater~in its manifest file under the \textless application\textgreater~tag \cite{Android:Net:Config}. On the positive side, only two apps were found to be susceptible to that vulnerability.

\subsection{CWEs}
\label{SS:CVEs}

This subsection details on all the high value CWEs that are pertinent to each app. From them, CWE-89 currently belongs to the list of top 25 most dangerous software weaknesses \cite{O:CWE:Mitre}, while CWE-295 and CWE-532 occupy positions 28 and 33 in the extended list, respectively.

\begin{itemize}
    
    
\item \emph{CWE-330}: The ``Use of insufficiently random values'' vulnerability pertains to the generation of predictable random values inside the app. This issue happens if the app uses an insecure random number generator. In OWASP top 10 mobile risks list (OWASP-10), this weakness is placed in the fifth position, namely ``insufficient cryptography''. Surprisingly, all but one app are potentially prone to this weakness.
    
\item \emph{CWE-276}: This CWE, namely ``Incorrect default permissions'', occurs if the app is granted unneeded read/write permissions. So, any affected file can be read/written from anyone. With reference to OWASP-10, this weakness is classified under M2, namely, ``insecure data storage''. As observed from the table, sixteen apps were potentially vulnerable to this weakness for at least one of the following reasons. The first, pertaining to the leftmost ``+'' sign, links to the fact that the app requests (read/write) access to the external storage. The second is related to the creation of a temp file, which may contain sensitive data. This is a major issue, since anyone can access folders that contain temp files, say, ``/data/local/tmp/*''.

\item \emph{CWE-532}: This weakness, namely, ``Insertion of sensitive information into log file'', emerges when a production app has enabled logging information to a file. While this feature may be of aid during the development stage of an app, it must be removed before the app becomes publicly available. In simple terms, an attacker could read these files and acquire any private information stored on them. All apps but two are potentially vulnerable to this issue.

\item \emph{CWE-312}: It is well-known as ``Cleartext storage of sensitive information'', and classified as M9 in OWASP-10. That is, when sensitive information, e.g., a username and/or password, are stored in cleartext form, anyone may be in position to read them. In some cases, this information may be stored inside the code of the app, e.g., in a configuration file. As observed from the table, seven apps were immune to this weakness.

\item \emph{CWE-89}: This hazardous weakness, titled ``Improper neutralization of special elements used in an SQL command ('SQL Injection')'' is classified as M7 in OWASP-10. It arises when the app does not sanitize or improperly sanitizes input stemming from an upstream component, say, from a Web form for user authentication. About two-thirds of the apps were found to be potentially vulnerable to this issue.
    
\item \emph{CWE-327}: It is referred to as ``Use of a broken or risky cryptographic algorithm'', and it belongs to M5 (``Insufficient Cryptography'') of OWASP-10. This weakness relates to the usage of obsolete or risky encryption or hash algorithms. As observed from the table, six apps may potentially use at least one obsolete hash algorithm, namely MD5 or SHA-1, and 3 of them support AES in ECB mode.

\item \emph{CWE-295}: This weakness titled ``Improper certificate validation'' is classified under M3 (``Insecure Communication'') in OWASP-10. This occurs in cases where the app is configured to trust an insecure or self-signed or any kind of certificate. This situation however may enable an opponent to mount man-in-the-middle (MitM) attacks. Five of the examined apps are potentially vulnerable to this weakness due to an insecure implementation of TLS.

\item \emph{CWE-749}: It is referred to as ``Exposed dangerous method or function'' and classified as M1 (``Improper Platform Usage'') in OWASP-10. This weakness is linked to a number of grave vulnerabilities, which each time depend on the underlying vulnerable function. Specifically, one app was found to offer an insecure \emph{WebView} implementation. The latter is used to display web content as part of an activity layout. In presence of this weakness, an attacker could possibly mount a MitM attack or even execute a Cross Site Scripting (XSS) injection. For more details regarding this issue, the interested reader may refer to the ``WebView'' section of \cite{O:Dev:Android:Security:Tips}.
    
\item \emph{CWE-919}: This weakness titled ``Weaknesses in Mobile Applications'' is directly related to CWE-749. Both of them address the same issue, but for another matter. For the examined apps, it was observed that only one of them has enabled the remote WebView debugging. That is, debug mode must be disabled before deploying a production application, otherwise anyone who can access an unlocked mobile device can easily obtain the app's data.


\end{itemize}

Altogether, the analysis demonstrated that a significant number of apps are susceptible to an array of high value common weaknesses. For instance, the great majority, i.e., 24 to 25 apps, were found to be prone to two of the revealed weaknesses, namely CWE-532 and CWE-330, respectively, while 18, 17, and 16 of them to CWE-312, CWE-89, and CWE-276, respectively.
Overall, as with many other software and hardware products, despite the fact that keeping up-to-date with common and high severity weaknesses aids in preventing security vulnerabilities and mitigating risk, this situation bespeaks a rather medium priority on security features.

\subsection{Tracker analysis}
\label{SS:Tracker:Analysis}

This subsection reports on third-party trackers that may be utilized by each app. Specifically, MobSF uses the open source \emph{Exodus-Privacy} \cite{O:Exodus:Privacy} webapp to analyze any detected tracker in the app's APK. We report only on two tracker categories which were spotted in the analyzed apps.

``Crash reporters'' focus on the crashes that may occur during the normal operation of the app. Upon a crash event, they send a notification message to the developers, informing them about the relevant error.
On the other hand, an ``analytics'' tracker collects any possible information regarding the usage of the app by the users, say, the time each user spent in the app, which features they used, and so forth.
For details about the third-party trackers issue in the mobile ecosystem, the interested reader can refer to the works in \cite{A:Trackers:Razaghpanah, A:Trackers:2:Vallina, A:Trackers3:Liu}.
    
App analysis revealed that 7 apps only employ \emph{Firebase} \cite{O:Firebase} or other Google analytics service as a means to measure users' engagement with them. On top of that, 17 apps were found to be free of any tracker, which certainly is on the plus side. For the rest two apps the following observations were made.


\emph{Microsoft Visual Studio App Center Crashes} \cite{O:Appcenter:Crashes} and \emph{Microsoft Visual Studio App Center Analytics} \cite{O:MS:Analytics} were detected in Smittestopp. The first tracker creates an automatic report, which includes any required information related to an app crash. When the user re-opens the app, the report is sent to the ``App Center''. In that regard, this tracker is categorized as crash reporter. Based on Exodus, the second tracker ``collects real-time analytics that highlight users' behavior. It also provides push notifications to mobile devices'', therefore it is categorized as an analytics one.
Lastly, \emph{Matomo} \cite{O:Matomo} and \emph{Bugsnag} \cite{O:Bugsnag} trackers were detected in CovTracer. The former is an open-source analytics tracker, while the latter a crash analytics one.

\subsection{Manifest analysis}
\label{SS:Manifest:Analysis}

Based on coarse-grained static analysis given in section \ref{S:High:Level:Static:Analysis}, we noted the use of several dangerous permissions from a privacy-invasive perspective. To deepen the analysis, we utilized MobSF to scrutinize the manifest file of each app, and possibly reveal any latent weaknesses. The focus here is on services, activities, and broadcast receivers. All of them, utilize intents and intent-filters. Based on the Android developer's guide, all of the aforementioned components, but the intents, must be declared in the manifest file of an app \cite{O:Android:Manifest}.

\emph{Intents} are basically message objects, which are used for either intra- or inter-app communication. They have three main usages, namely start an activity, initiate a service, and deliver a broadcast. There are two types of intents, namely explicit and implicit. The former is used to handle messages within the app, while the latter to transfer messages towards another capable app. On the other hand, an \emph{intent filter} is an expression in an app's manifest file that determines the kind of intents the component would wish to obtain. In this respect, intent-filters are responsible for handling any implicit intent, e.g., broadcast receivers, and capturing system-oriented broadcast messages, where the specific broadcast message is contained inside an intent. This means that for an app to receive such a broadcast message, it must declare in its manifest file a matching intent-filter.

The \emph{service} app component can perform operations without needing a user interface (UI), e.g., transferring files from one app to another without involving the user. There are three types of services; \emph{foreground}, which is discernible to the user, \emph{background}, which is not directly perceivable by the user, and \emph{bound}, which is used to bound a specific service with an app.
Lastly, \emph{activities} is an key component of any Android app. An activity, comprises a single, specific thing the end-user can perform, and it usually involves a UI. For example, when a user opens an app, the \emph{main activity} is typically executed.

App analysis revealed that some of these three types of components did not declare a permission in the respective manifest file. Namely, according to Android developer security tips \cite{O:Dev:Android:Security:Tips}, when a component is declared in the manifest file is by default enabled to communicate with other apps. To restrict this functionality, the developer must declare among others a permission to this component. Next, any other app must possess the same permission for being able to communicate with this component. Note that this is a minimum security measure, meaning it is not adequate to fully secure the component, but only lessen the magnitude of the issue. Therefore, the omission to at least provide the right permission to such a component is identified as a high severity weakness.

Precisely, the results obtained per app are summarized in table \ref{T:Manifest:Components}. The left part of the table divides the components of interest in two categories, namely intent-filter on and off. The first (on) means that the app can send and also receive intents which target that specific component. When the received intent contains a malicious content, it can potentially compromise that app, e.g., bypass authentication \cite{O:Intent:Malicious}. The second (off) signifies that the app can potentially only send any information that will be requested from another app. Meaning that it is possible for the app to unwillingly leak sensitive information to an assailant. From the table, it is inferred that in either of these two categories and across all the three components, a considerably but generally not large number of apps, i.e., 1, 15, 3 and 8, 3, and 2, respectively neglect to declare the appropriate permissions.

\begin{table}[htpb]
\begin{adjustbox}{width=1\textwidth}
\begin{tabular}{|l|l|l|l|l|l|l|l|}
\hline
\multicolumn{1}{|c|}{\multirow{2}{*}{\textbf{App Name}}} &
  \multicolumn{3}{|c|}{\textbf{Intent-filter on}} &
  \multicolumn{3}{|c|}{\textbf{Intent-filter off}} &
  \multicolumn{1}{|c|}{\multirow{2}{*}{\textbf{Cleartext}}} \\ \cline{2-7}
\multicolumn{1}{|c|}{} &
  \multicolumn{1}{|c|}{\textbf{Service}} &
  \multicolumn{1}{|c|}{\textbf{Broadcast Receiver}} &
  \multicolumn{1}{|c|}{\textbf{Activity}} &
  \multicolumn{1}{|c|}{\textbf{Service}} &
  \multicolumn{1}{|c|}{\textbf{Broadcast Receiver}} &
  \multicolumn{1}{|c|}{\textbf{Activity}} &
  \multicolumn{1}{|c|}{} \\ \hline
Stopp Corona  & -- & -- & 1 & -- & 1 &-- &--\\
Coronalert  & -- & -- & -- & 1 & -- & -- &--\\
ViruSafe  & -- & -- & -- & -- & 1 & --&-- \\
Stop COVID-19 & -- & 1 & -- & -- & -- & -- &--\\
CovTracer & -- & -- & -- & -- & 1 & 1&--\\
eRouska & -- & -- & -- & 1 & -- & 2 &--\\
Smittestop & -- & 4 & 1 & -- & -- & -- &--\\
Hoia & -- & 1 & -- & -- & -- & -- &--\\
Koronavilkku & -- & -- & -- & -- &--  & -- &--\\
TousAntiCovid & -- & -- & -- & -- &--  & --&--\\
Corona-Warn-App & -- & -- & -- & 1 & -- & -- &--\\
VirusRadar & -- & -- & -- & -- &--  & -- &--\\
COVID Tracker & -- & 1 & -- & -- & -- & -- &--\\
Immuni & -- & 1 & -- & -- & -- & -- &--\\
Apturi Covid & -- & 1 & -- & -- & -- & -- &--\\
Korona Stop LT & -- & -- & -- & 1 & -- & -- &--\\
COVID Alert & -- & 1 & -- & -- & -- & -- &--\\
CoronaMelder & -- & 2 & -- & 1 & -- & -- &--\\
Smittestopp & -- & 3 & 1 & -- & -- & -- &--\\
STOP COVID - ProteGO Safe &--  & 1 & -- & 1 & -- & --&--\\
STAYAWAY COVID & -- & 1 & -- & -- & -- & -- &--\\
ZostanZdravy & 1 & 1 & -- & 1 & -- & -- & + \\
\#OstaniZdrav & -- & -- & -- & 1 & -- & -- &--\\
Radar COVID & -- & 1 & -- & -- & -- & --&--\\
SwissCovid & -- & 1 & -- & -- & -- & -- &--\\ 
NHS COVID-19 & -- & 2 & -- & 1 & -- & -- &--\\ \hline 
Total & 1 & 15 & 3 & 8 & 3 & 2 & 1\\ \hline
\end{tabular}
\end{adjustbox}
\caption{Potentially vulnerable components in the manifest file of each app.}

\label{T:Manifest:Components}
\end{table}

The last column of table \ref{T:Manifest:Components} exhibits information on another important feature related to the contents of the manifest file. This is the \emph{android:usesCleartextTraffic} flag in the \textless application\textgreater~manifest element, which designates whether the app intends to use cleartext network traffic, including cleartext HTTP. As seen from the table, only one app allows this functionality. Naturally, this may compromise the privacy of the end-user.

\subsection{Shared library analysis}
\label{SS:Shared:Library:Binary:Analysis}

This type of analysis pertains to the shared libraries (with the extension \emph{.so}) an app may incorporate. These libraries, are typically written in C and compiled with the Android native development kit (NDK) toolset \cite{O:Dev:Android:NDK}. Android capitalizes on this logic for achieving better performance and reusing existing C libraries, without translating them to Java. These libraries are loaded into memory at runtime. Bear in mind that security and privacy issues with the use of shared libraries have been already identified in the Android literature \cite{A:Shared:Libraries:Tegawend, A:Shared:Libraries:Taylor, Backes-2016}.

To investigate if and to what degree this issue applies to the examined apps, we present the number of potentially vulnerable shared libraries per app in table \ref{T:Shared:Libraries}. Again, we only consider high severity potential vulnerabilities, which pertain to four exploit mitigation techniques, namely, no-execute (NX), position-independent executable (PIE), Stack Canary, and relocation read-only (RELRO). These countermeasures, more accurately referred to as memory corruption mitigation techniques, are specific to C language, and if neglected may create space for memory-based exploits, which inevitably migrate to the affected Android app. As seen from the table, about half of the analyzed apps were found to incorporate shared libraries that overlook two of the above-mentioned remedies, namely PIE and RELRO, while only two apps have a plethora of libraries that disregard all four exploit mitigation techniques.

\begin{table}[htpb]
\begin{tabular}{|l|r|r|r|r|}
\hline
\multicolumn{1}{|l|}{\textbf{App Name}} &
  \multicolumn{1}{|c|}{\textbf{NX}} &
  \multicolumn{1}{|c|}{\textbf{PIE}} &
  \multicolumn{1}{|c|}{\textbf{Stack Canary}} &
  \multicolumn{1}{|c|}{\textbf{RELRO}} \\ \hline
Stopp Corona  & --  & --  & --  & --  \\
Coronalert    & --  & 8   & --  & 8   \\
ViruSafe      & --  & --  & --  & --  \\
Stop COVID-19 & --  & --  & --  & --  \\
CovTracer     & --  & 92  & --  & 92  \\
eRouska       & --  & --  & --  & --  \\
Smittestop    & 409 & 560 & 546 & 568 \\
Hoia          & --  & --  & --  & --  \\
Koronavilkku  & --  & 4   & --  & 4   \\
TousAntiCovid & --  & --  & --  & --  \\
Corona-Warn-App   & --  & 8   & --  & 8   \\
VirusRadar    & --  & --  & --  & --  \\
COVID Tracker & --  & 83  & --  & 83  \\
Immuni        & --  & 4   & --  & 4   \\
Apturi Covid  & --  & --  & --  & --  \\
Korona Stop LT& --  & 8   & --  & 8   \\
COVID Alert    & --  & --  & --  & --  \\
CoronaMelder  & --  & --  & --  & --  \\
Smittestopp   & 525 & 594 & 583 & 611 \\
STOP COVID - ProteGO Safe& --  & 4   & --  & 4   \\
STAYAWAY COVID & --  & 80  & --  & 80  \\
ZostanZdravy & --  & --  & --  & --  \\
\#OstaniZdrav & --  & 8   & --  & 8   \\
Radar COVID   & --  & --  & --  & --  \\
SwissCovid    & --  & --  & --  & --  \\
NHS COVID-19  & --  & --  & --  & -- \\ \hline
Total         & 2 & 12 & 2 & 12 \\ \hline
\end{tabular}
\label{T:Shared:Libraries}
\caption{Shared library issues per app.}

\end{table}

An OS that endorses the NX bit - a feature of the memory management unit of some CPUs - may tag specific sections of the memory as non-executable, meaning that the CPU will deny to execute any code residing in that region. If the NX bit is not set on the library, an attacker may be able to mount a buffer overflow. That is, frequently, such attacks place code in a program's data region or stack, and subsequently jump to it. But if all writable addresses are non-executable (through the \emph{-z noexecstack} compiler flag), such an attack is blocked. As observed from the table, a couple of apps were found to incorporate libraries that allow for executable writable addresses in memory.

PIEs on the other hand are executable binaries which are made from position-independent codes (PICs). The latter are used by shared libraries for loading their code into memory at runtime, without overlapping with other shared libraries that already exist there. Actually, this is a common mechanism to harden Executable and Linkable Format (ELF) binaries.
As seen from the table, about half of the apps incorporate shared libraries which were built without enabling the position independent code flag (\emph{-f PIC}). This could be exploited by an attacker, forcing the app to jump to a specific part of the memory that contains malicious code.

The use of Stack Canaries is a well-known defense against memory corruption attacks; it is a value placed on the stack with the intention to be overwritten by a stack buffer that overflows to the return address. If this protection is withheld, the app is prone to stack buffer overflow attacks, say, those that aim to overwrite certain parts of memory with malicious code. Our results show that only two apps embrace shared libraries which neglect this defence.

Relocation Read-Only (RELRO) is another mechanism to harden ELF binaries by rendering some binary sections read-only. Precisely, RELRO ensures that the global offset table (GOT) - a lookup table used by a dynamically linked ELF binary to dynamically resolve functions that are located in shared libraries - cannot be overwritten. For the interested reader, an example of a RELPO-relevant attack is given in \cite{O:Shared:Libraries:RELRO}.
With reference to table \ref{T:Shared:Libraries}, at least four libraries in roughly half of the apps do not make use of the RELRO defence.



\subsection{Outdated software components analysis}
\label{SS:Third:Party:Libraries}

For both the desktop and mobile platforms, third-party components, say, libraries, comprise one of the keystones of modern software development. However, as already mentioned in subsection \ref{SS:Shared:Library:Binary:Analysis}, the benefit of reusing third-party code may be largely neutralized, if that code is buggy or outdated. This may significantly augment the attack surface of the app and expose end-users to security and privacy risks stemming from those external software components. Indeed, the importance of updatability of such Libraries on the Android platform has been frequently pinpointed in the literature \cite{A:Third:Party:Libraries:Derr:2017, A:Third:Party:Libraries:Salza:2018}. Put simply, it has been shown that many Android apps do not update their third-party libraries, remaining vulnerable to a range of CVEs.

As already mentioned, to delve into this issue under the perspective of the examined apps, we employed the Ostorlab tool. The outcomes of this type of analysis per app are summarized in table \ref{T:CVE:Analysis}. As observed from the table, approximately the one-third of the apps (8) make use of one at least outdated library. However, as already stated, such a shortcoming is tightly connected to one or more CVEs, meaning that the respective app is susceptible to publicly disclosed security flaws. In the following, we briefly describe such issues per shared library by just enumerating the relevant CVEs. The interested reader may in addition consult the respective CVE page in \cite{O:nvd:nist}.

\begin{table}[htpb]
\centering
\begin{adjustbox}{width=0.7\textwidth}
\begin{tabular}{|l|c|c|c|}
\hline
\textbf{App Name} &
\multicolumn{1}{l|}{\textbf{SQLite}} &
\multicolumn{1}{l|}{\textbf{OpenSSL}} &
\textbf{libjpeg} \\ \hline
    Stopp Corona &  -- & -- &  --\\
    Coronalert &  + & +  & -- \\
    ViruSafe & --  & --  & -- \\
    Stop COVID-19 &-- & --  & --\\
    CovTracer &  -- &  -- & +\\
    eRouska & --  &--& -- \\
    Smittestop &  -- & --  &  --\\
    Hoia &-- & --  & -- \\
    Koronavikku & +  & +  & -- \\
    TousAntiCovid &-- & --  & -- \\
    Corona-Warn-App & +  & +  & -- \\
    VirusRadar &  -- & --  &--  \\
    COVID Tracker & +  &  + & + \\
    Immuni &-- & --  & -- \\
    Apturi Covid &  -- &  -- &  --\\
    Korona Stop LT &  + & +  & -- \\
    COVID Alert &  -- & --  & -- \\
    CoronaMelder &-- & --  & -- \\
    Smittestopp & --  &--&  --\\
    STOP COVID - ProteGO Safe &  -- & --  &--  \\
    STAYAWAY COVID & --  & --  &+\\
    ZostanZdravy & --  & --  & -- \\
    \#OstaniZdrav & +  & +  & -- \\
    Radar COVID &-- & --  & -- \\
    SwissCovid &  -- &  -- & -- \\
    NHS COVID-19 &  -- & --  & -- \\\hline
    Total & 6 & 6 & 3 \\\hline
\end{tabular}
\end{adjustbox}
\caption{Outdated third-party software components per app.}

\label{T:CVE:Analysis}
\end{table}

Specifically, table \ref{T:CVE:Analysis} reveals that the same 6 apps utilize an outdated version of the well-known \emph{SQLite} \cite{O:sqlite} and \emph{OpenSSL} \cite{O:openssl} libraries, v.3.31.0 and v.1.1.1g, respectively. As it is well-known, the first library, implements a lightweight, full-featured, SQL database engine, while the second provides an open-source implementation of the SSL and TLS protocols. With reference to the SQLite documentation \cite{O:sqlite}, obsolete versions of SQLite may be vulnerable to denial of service (DoS) attacks. Some pertinent CVEs to this case are CVE-2020-13632, CVE-2020-13434, and CVE-2020-13435. Even worse, several of them, say, CVE-2020-11655, CVE-2020-11656, and CVE-2020-13632 are classified as being of high or critical severity. With respect to OpenSSL, along with DoS-related CVEs, we observed 3 others of medium severity, namely CVE-2018-12433, CVE-2018-12438, and CVE-2020-1971. Clearly, these kind of vulnerabilities can be potentially weaponized for breaking the encryption between the app and the server.

Lastly, three apps were found to incorporate an obsolete version (v1.5.3) of the \emph{libjpeg} \cite{O:libjpeg} library, used to read and write JPEG image files. With reference to CVE-2018-14498 and CVE-2018-20330, this version may be vulnerable to DoS attacks.


\subsection{Taint analysis}
\label{SS:Taint:Analysis}

Static taint analysis, i.e., a form of information flow analysis, has been performed with the aid of the Ostorlab tool. Generally, this type of examination can identify data leakage type of problems in the examined code. This typically refers to a variety of user or other kind of input sanitization hiccups, that may facilitate intent injection, SQL injection, password tracking, or even buffer overflows. To do so, taint analysis uses a script, which tags every private data of interest, known as the \emph{source}. By following each source throughout the code as a flow, the analysis may reveal every code snippet that potentially has a leakage, the so-called \emph{sink}. A notable remark here is that taint analysis may yield a considerable number of false positives. Therefore, to reduce the error factor, the results of this subsection have been cross-checked with those of manifest analysis given in subsection \ref{SS:Manifest:Analysis}, and we preserved only the intersection of the two sets based on the findings of taint analysis. Overall, an approximately 15\% of the taint analysis results were found to be common between the two sets.

\begin{figure}[htpb]
    \centering
    \includegraphics[width=1\textwidth]{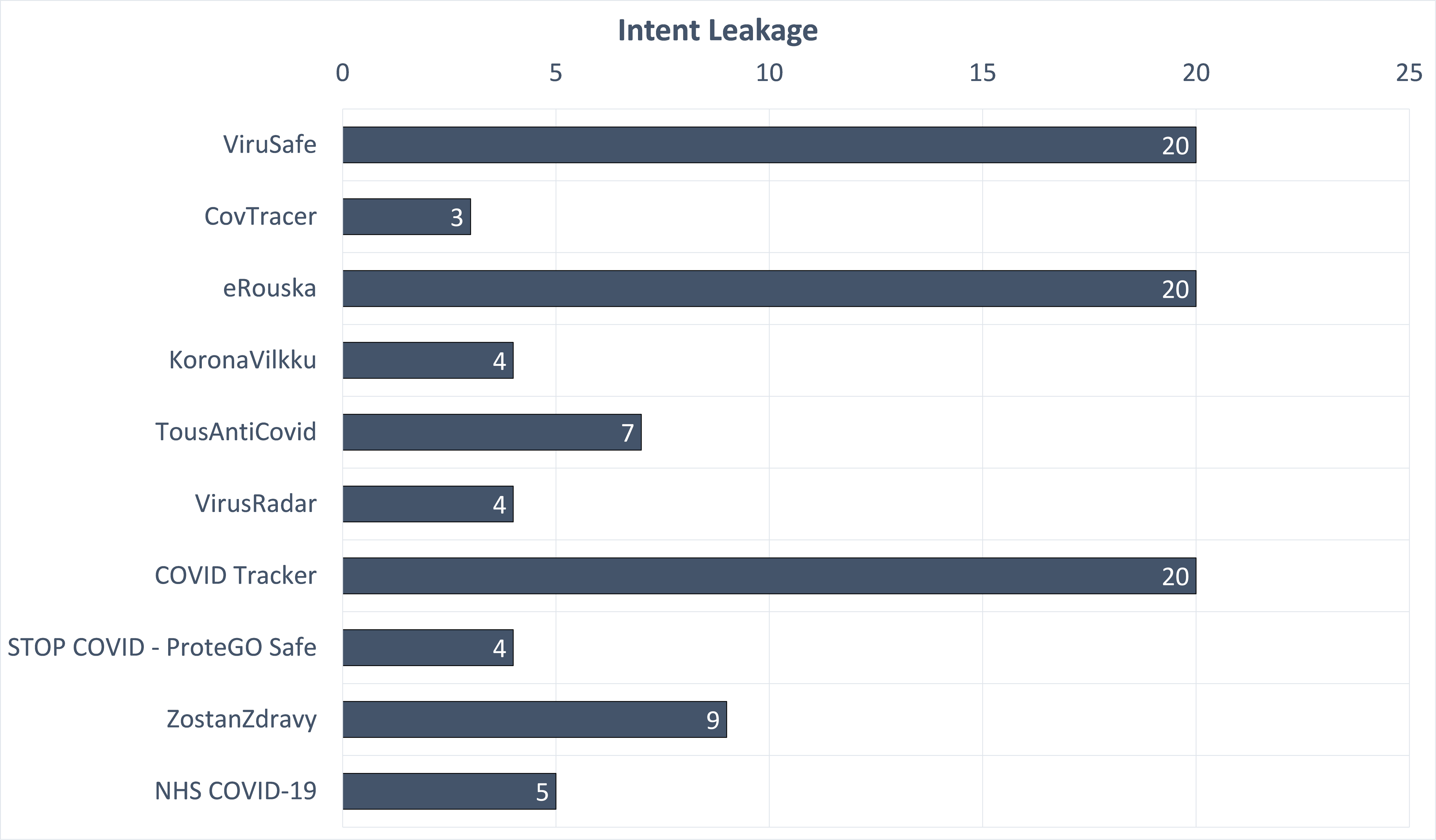}
    \caption{Issues identified through taint analysis.}
    \label{F:Taint:Analysis}
\end{figure}

For the examined apps, taint analysis only identified issues related to intent leakage. As already mentioned in subsection \ref{SS:Manifest:Analysis}, this type of leakage is related to three major components of the Android OS; broadcast receivers, services, and activities. Basically, properly sanitizing these components in the manifest file, means less leaked intents. More precisely, the results of applying taint analysis to each of the available apps are illustrated in figure \ref{F:Taint:Analysis}. The number inside each bar designates the quantity of the issues per app. As observed, potential issues have been identified for less than half, i.e., 10, of the examined apps; seven of them showed a relatively low number of issues, while the rest a considerably larger number.


\section{Dynamic analysis}
\label{S:Dynamic:Analysis}


As already mentioned in section \ref{S:methodology}, the features collected during this phase pertain to Java classes and network traffic. The instrumentation code fed to Androtomist is given in code Listing \ref{lst:code2}, where the tracked classes are defined in the \emph{objectsToLookFor} variable. Precisely, during runtime, the code logs specific Java classes related to network traffic, location, mobile network, and Bluetooth, according to the API documentation, as follows:

\begin{itemize}

\item Network traffic: java.net.URLConnection, java.net.URL, java.net.Socket, including all relevant parameters, that is, URLs, IP addresses, and network protocols.

\item Location: android.location.Location, android.location.LocationManager, and android.location.LocationProvider. An object of the first class represents a geographic location, including latitude, longitude, and timestamp. The second class provides access to the system's location services, while the location provider abstract class can be implemented to periodically report on the geographical location of the device. 


\item Bluetooth: android.bluetooth.BluetoothAdapter, android.bluetooth.BluetoothManager, android.bluetooth.BluetoothDevice, android.bluetooth.BluetoothClass.

\item Mobile network: android.telephony.TelephonyManager. This class is of interest because it provides access to pieces of data regarding the telephony services on the mobile device. Precisely, the methods of this class can be called by an app to learn the available telephony services and states, and obtain certain types of sensitive subscriber information, including the unique subscriber identity, i.e., the International Mobile Subscriber Identity (IMSI).

\end{itemize}

\begin{lstlisting}[language=Java, caption = The instrumentation code exploited to hook Java classes,label={lst:code2}]
//Objects to hook
var objectsToLookFor = ["android.location.Location",
"android.location.LocationManager",
"android.location.LocationProvider",
"java.net.Socket","java.net.URLConnection",
"java.net.URL",
"android.telephony.TelephonyManager",
"android.bluetooth.BluetoothAdapter",
"android.bluetooth.BluetoothManager",
"android.bluetooth.BluetoothDevice",
"android.bluetooth.BluetoothClass",
"android.hardware.Camera",
"android.hardware.camera2.CameraManager",
"dalvik.system.DexClassLoader"];

for (var i in objectsToLookFor) {
	Java.perform(function () {
		Java.choose(objectsToLookFor[i], {
			"onMatch": function (instance) {
			    console.log("\nProcess_has_Instantiated_instance_of: " + objectsToLookFor[i]);
				console.log("Details: " + instance.toString());
			},
			"onComplete": function () {
			}
		});
	});
}

\end{lstlisting}

The results produced by dynamic analysis are recapitulated in table \ref{T:dynamicanalysis}. The four rightmost columns of the table
designate if the examined app has (a) produced network traffic, (b) instantiated the Java class android.location.LocationManager, (c) instantiated the Java class android.telephony.TelephonyManager, and (d) instantiated either or both the Java classes android.bluetooth.BluetoothAdapter and android.bluetooth.BluetoothManager, respectively. As seen in the table, two apps asked for phone number authentication through a verification code received via SMS to properly work. So, given that the Nvidia Shield K tablet device is SIM-less, these apps were excluded from the current analysis. All in all, the following key takeaways emerge.

\begin{itemize}

\item All the examined apps but three produced network traffic, meaning they visited specific URLs as detailed in subsection \ref{SS:Analysis:of:apps}.

\item Thirteen apps employed the android.location.LocationManager class, which provides access to the system's location services and allows apps to obtain periodic updates of the device's geographical location. Actually, this result confirms the outcomes of subsection \ref{SS:Analysis:of:apps} with reference to the \seqsplit{android/location/LocationManager} API call.

\item All the examined apps used the android.telephony.TelephonyManager. This outcome comes in addition to that of subsection \ref{SS:Analysis:of:apps} in relation to the \seqsplit{android/telephony/TelephonyManager} API call.

\item All the examined apps but one instantiated either or both the android.bluetooth.BluetoothAdapter and android.bluetooth.BluetoothManager Java classes. The first class lets the app conduct basic Bluetooth tasks, including starting up device discovery, while the latter offers a high-level manager employed to get an instance of an BluetoothAdapter and to perform overall Bluetooth management.

\end{itemize}

\begin{table}[hbt!]
\centering
\footnotesize
\begin{tabular}{|p{4cm}|c|c|c|c|}
\hline
\textbf{App} & \textbf{Net. traffic} & \textbf{Location} & \textbf{Telephony} & \textbf{Bluetooth}\\
\hline
Stopp Corona        & + & -- & + & + \\
Coronalert          & + & + & + & + \\
ViruSafe*           & N/A & N/A & N/A & N/A \\
Stop COVID-19       & + & + & + & + \\
CovTracer           & + & + & + & + \\
eRouska             & + & + & + & + \\
Smittestop          & -- & -- & + & + \\
Hoia                & + & -- & + & + \\
KoronaVilkku        & -- & + & + & + \\
TousAntiCovid       & + & -- & + & + \\
Corona-Warn-App     & + & + & + & + \\
VirusRadar*         & N/A & N/A & N/A & N/A \\
COVID Tracker       & + & + & + & + \\
Immuni              & + & + & + & + \\
Apturi Covid        & + & + & + & + \\
Korona Stop LT      & + & + & + & + \\
COVID Alert          & + & -- & + & + \\
CoronaMelder        & + & + & + & + \\
Smittestopp         & -- & -- & + & + \\
STOP COVID - ProteGO Safe & + & + & + & +\\
STAYAWAY COVID      & + & -- & + & + \\
ZostanZdravy        & + & -- & + & -- \\
\#OstaniZdrav       & + & + & + & + \\
Radar COVID         & + & -- & + & + \\
SwissCovid          & + & -- & + & + \\
NHS COVID-19        & + & -- & + & + \\
\hline
Total               & 21 & 13 & 24 & 23 \\
\hline
\end{tabular}
\caption{Dynamic analysis results. *Excluded from analysis.}

\label{T:dynamicanalysis} 
\end{table}

Besides the above-mentioned general remarks, the following observations were done after comparing the results of table \ref{T:dynamicanalysis} with those from Tables \ref{T:permissions:baseline} and \ref{T:apicalls:baseline}:

\begin{itemize}

\item As shown in table \ref{T:apicalls:baseline}, according to the findings of static analysis, all apps included API calls related to location, while in dynamic analysis only 13 of them did instantiate relevant classes. As previously mentioned, according to the API, all location methods require either or both the ACCESS\_COARSE\_LOCATION and ACCESS\_FINE\_LOCATION permissions. As a result, \emph{ViruSafe}, \emph{TousAntiCovid}, \emph{VirusRadar}, and \emph{ZostanZdravy}, namely, all the not GAEN-based apps, do declare the necessary permissions, and thus are indeed able to get the device's location data.
    
\item With reference to the results of subsection \ref{SS:Analysis:of:apps}, the \emph{ZostanZdravy} app did not instantiate any Bluetooth classes.
    
\item As already mentioned, all apps instantiated the \seqsplit{android.telephony.TelephonyManager} class, which pertains to the cellular network. Nevertheless, according to static analysis results, only 3 apps included API calls related to the same category.

\end{itemize}

\section{Conclusions}
\label{S:Conclusions}

The merit of mobile digital contact tracing apps in swiftly and accurately identifying and alerting individuals at risk in large-scale is rather unquestionable. Nevertheless, along with other considerations, for such a privacy-sensitive app to gain widespread adoption, certain security and data protection requirements need to be met. Actually, this reluctance can be perceived from the information contained in Google Play Store, where it seems that (i) based on the app downloads, the penetration of most apps in the population is still not very high, and (ii) the rather fair average evaluation score they receive can be translated as an indication of undesired, unneeded, or sometimes buggy or problematic functionality.

The empirical results reported in this work can serve a triad of goals. The first, targets app designers and developers, and relates to strengthening the sound design of upcoming versions of these apps, as well as future contact tracing apps, in terms of security and end-user privacy. Also, given that the findings of this endeavor reflect the general proclivity in this ecosystem, it can be straightforwardly used as a basis for further research, focusing on promoting best practices that facilitate the minimization of the attack surface, obstruct data leakage, and nurture security and privacy by design overall. Especially for privacy, the root issue being at stake here is the so-called principle of minimal privilege, mandating that a user, process, or program, must be by default enabled to only access the data and resources that are necessary for fulfilling its mission. Last but not least, our findings can aid policy makers in shaping a better understanding of this ecosystem. That is, in this ``black swan'' situation, policy makers must be provided with sufficient and suitable data and knowledge to formulate strategies for promoting mass digital tracing app acceptance, thus contributing to the greater good. This would augment the chances of favorable app exploitation. Under this prism, policy makers need to not only be aware of which app characteristics matter most for, say, the reluctant or opposed citizens, but also be informed of the current situation and if and to what degree it coincides with the achievement of objectives and priorities in the mid- or long-run.

In this context, this work attempts to answer two rudimentary, but decisive questions: Do the currently deployed digital contact apps (a) reduce their functionality to the bare minimum?, and (b) remain free of known misconfigurations, weaknesses, and vulnerabilities? If so, user privacy is well-preserved, data protection is ensured, and the app's attack surface is decreased. To answer both these matters in a tangible way, we meticulously analyzed all official mobile contact tracing apps currently deployed by European countries.

The results stemming for static analysis lead to a number of conclusions: (a) All GAEN-based apps respect the minimum set of required permissions, but exceptions do apply; two of them demand in addition access to the camera, say, for enabling the user to scan a QR code, (b) a limited number of GAEN-incompatible apps request at least a couple of extra sensitive additional permissions, (c) at least $\approx$62\% of the apps were found to be potentially exposed to 5 out of the total 9 high value CWEs. This percentage increases to about 92\% and 96\% for CWE-532 and CWE-330, respectively, (d) roughly 88\% and 46\% of the apps exhibited more than one issue in their manifest file and shared libraries, respectively, (e) outdated software component exploration and taint analysis revealed that approximately one-quarter and one-third of the apps, respectively, have at least one issue, (f) on the plus side, almost 65\% of the apps were found to be free of any third-party tracker, and only 2 apps were found to incorporate at least one third-party tracker. 

On the other hand, dynamic analysis also yielded significant outcomes, allowing us to verify the results stemming from static analysis. Overall, it was perceived that in most cases, the outcomes coincide, but exceptions do exist. For instance, despite that all apps contain at least one location-related API call, about half (13) of them actually instantiate the relevant classes. Interestingly, the same observation stands true in the opposite direction for telephony-related API calls. Namely, with reference to static analysis, only 3 apps were found to include such calls, while dynamic analysis showed that all apps instantiated the relevant class. In any case, dynamic analysis is further to be addressed in a future work by hooking additional operations, i.e., refining the instrumentation code, and via the use of control flow graphs.

Finally yet importantly, an app may include code (logged in static or triggered during dynamic analysis) for which it has not asked permission in the manifest file. Typically, in such a case, the app cannot use these pieces of code, but this situation does not apply to rooted phones. That is, any permission listed in the manifest file provides access to some of the data and functions offered by the Android OS, but root privileges grant access to data and functions that were never intended to be shared or used. Not to mention, of course, the fact that an app with root access can automatically configure all permissions as will. In any case, deliberately rooting or jailbreaking a smartphone, drastically alters the security posture of the device, entailing the circumvention of the native security restrictions put in place by the operating system. This may expose data and apps to numerous threats, and thus this practice is exclusively under the end-user's own responsibility.


\begin{thebibliography}{10}

\bibitem{Martin}
Martin T, Karopoulos G, Ramos JLH, Kambourakis G, Fovino IN.
\newblock Demystifying COVID-19 Digital Contact Tracing: A Survey on Frameworks
  and Mobile Apps.
\newblock Wirel Commun Mob Comput. 2020;2020:8851429:1--8851429:29.
\newblock doi:{10.1155/2020/8851429}.

\bibitem{GoogleApplecontacttracing}
{Apple}. Privacy-Preserving Contact Tracing;.
\newblock \url{https://www.apple.com/covid19/contacttracing}.

\bibitem{Restricted-API}
{Apple developer forums}. How to enable Exposure Notification capability?;.
\newblock \url{https://developer.apple.com/forums/thread/132591}.

\bibitem{XDA-2021}
{XDA}. Countries using Google and Apple’s COVID-19 Contact Tracing API;.
\newblock
  \url{https://www.xda-developers.com/google-apple-covid-19-contact-tracing-exposure-notifications-api-app-list-countries/}.

\bibitem{DP-3T}
{DP-3T}. Decentralized Privacy-Preserving Proximity Tracing;.
\newblock \url{https://github.com/DP-3T}.

\bibitem{PEPP-PT}
{PEPP-PT, GitHub}. Pan-European Privacy-Preserving Proximity Tracing;.
\newblock \url{https://github.com/pepp-pt}.

\bibitem{Fraunhofer}
AISEC F. Pandemic Contact Tracing Apps: DP-3T, PEPP-PT NTK, and ROBERT from a
  Privacy Perspective; 2020.
\newblock Cryptology ePrint Archive, Report 2020/489.

\bibitem{Avitabile}
Avitabile G, Botta V, Iovino V, Visconti I. Towards Defeating Mass Surveillance
  and SARS-CoV-2: The Pronto-C2 Fully Decentralized Automatic Contact Tracing
  System; 2020.
\newblock Cryptology ePrint Archive, Report 2020/493.

\bibitem{cho2020contact}
Cho H, Ippolito D, Yu YW. Contact Tracing Mobile Apps for COVID-19: Privacy
  Considerations and Related Trade-offs; 2020.
\newblock arXiv 2003.11511v2.

\bibitem{Raskar}
et~al RR. Apps Gone Rogue: Maintaining Personal Privacy in an Epidemic; 2020.
\newblock arXiv 2003.08567v1.

\bibitem{Nadeem}
Ahmed N, Michelin RA, Xue W, Ruj S, Malaney R, Kanhere SS, et~al.. A Survey of
  COVID-19 Contact Tracing Apps; 2020.
\newblock arXiV 2006.10306v2.

\bibitem{Jinfeng}
Li J, Guo X. COVID-19 Contact-tracing Apps: a Survey on the Global Deployment
  and Challenges; 2020.
\newblock arXiv 2005.03599.

\bibitem{samhi-2021}
Samhi J, Allix K, Bissyandé TF, Klein J. A First Look at Android Applications
  in Google Play related to Covid-19; 2021.
\newblock arXiv 2006.11002.

\bibitem{Trang-2020}
Trang S, Trenz M, Weiger WH, Tarafdar M, Cheung CMK.
\newblock One app to trace them all? Examining app specifications for mass
  acceptance of contact-tracing apps.
\newblock Eur J Inf Syst. 2020;29(4):415--428.
\newblock doi:{10.1080/0960085X.2020.1784046}.

\bibitem{O:Stopp:Corona}
{Google Play}. Stopp Corona;.
\newblock
  \url{https://play.google.com/store/apps/details?id=at.roteskreuz.stopcorona}.

\bibitem{O:Coronalert}
{Google Play}. Coronalert;.
\newblock
  \url{https://play.google.com/store/apps/details?id=be.sciensano.coronalert}.

\bibitem{O:Virusafe}
{Google Play}. ViruSafe;.
\newblock
  \url{https://play.google.com/store/apps/details?id=bg.government.virusafe}.

\bibitem{O:Stop:COVID:19}
{Google Play}. Stop COVID-19;.
\newblock
  \url{https://play.google.com/store/apps/details?id=hr.miz.evidencijakontakata}.

\bibitem{Cyprus:New}
{Google Play}. CovTracer-EN;.
\newblock
  \url{https://play.google.com/store/apps/details?id=cy.gov.dmrid.covtracer&hl=en_US&gl=US}.

\bibitem{O:eRouska}
{Google Play}. eRouska;.
\newblock
  \url{https://play.google.com/store/apps/details?id=cz.covid19cz.erouska}.

\bibitem{O:Smittestop}
{Google Play}. Smittestop;.
\newblock
  \url{https://play.google.com/store/apps/details?id=com.netcompany.smittestop_exposure_notification}.

\bibitem{O:Hoia}
{Google Play}. Hoia;.
\newblock \url{https://play.google.com/store/apps/details?id=ee.tehik.hoia}.

\bibitem{O:Koronavilkku}
{Google Play}. Koronavilkku;.
\newblock
  \url{https://play.google.com/store/apps/details?id=fi.thl.koronahaavi}.

\bibitem{O:TousAntiCovid}
{Google Play}. TousAntiCovid;.
\newblock
  \url{https://play.google.com/store/apps/details?id=fr.gouv.android.stopcovid}.

\bibitem{O:Corona:Warn:App}
{Google Play}. Corona-Warn-App;.
\newblock
  \url{https://play.google.com/store/apps/details?id=de.rki.coronawarnapp}.

\bibitem{O:VirusRadar}
{Google Play}. VirusRadar;.
\newblock
  \url{https://play.google.com/store/apps/details?id=hu.gov.virusradar}.

\bibitem{O:COVID:Tracker}
{Google Play}. COVID Tracker;.
\newblock
  \url{https://play.google.com/store/apps/details?id=com.covidtracker.hse}.

\bibitem{O:Immuni}
{Google Play}. Immuni;.
\newblock
  \url{https://play.google.com/store/apps/details?id=it.ministerodellasalute.immuni}.

\bibitem{O:Apturi:Covid}
{Google Play}. Apturi Covid;.
\newblock
  \url{https://play.google.com/store/apps/details?id=it.ministerodellasalute.immuni}.

\bibitem{O:Korona:Stop:LT}
{Google Play}. Korona Stop LT;.
\newblock
  \url{https://play.google.com/store/apps/details?id=lt.nvsc.coronawarnapp}.

\bibitem{O:COVIDAlert}
{Google Play}. COVIDAlert;.
\newblock \url{https://play.google.com/store/apps/details?id=mt.gov.dp3t}.

\bibitem{O:CoronaMelder}
{Google Play}. CoronaMelder;.
\newblock
  \url{https://play.google.com/store/apps/details?id=nl.rijksoverheid.en}.

\bibitem{O:Smittestopp}
{Google Play}. Smittestopp;.
\newblock
  \url{https://play.google.com/store/apps/details?id=no.fhi.smittestopp_exposure_notification}.

\bibitem{O:STOP:COVID:ProteGO}
{Google Play}. STOP COVID - ProteGO Safe;.
\newblock
  \url{https://play.google.com/store/apps/details?id=pl.gov.mc.protegosafe}.

\bibitem{O:STAYAWAY:COVID}
{Google Play}. STAYAWAY COVID;.
\newblock
  \url{https://play.google.com/store/apps/details?id=fct.inesctec.stayaway}.

\bibitem{O:ZostanZdravy}
{Google Play}. ZostanZdravy;.
\newblock \url{https://www.old.korona.gov.sk/en/COVID19-ZostanZdravy.php}.

\bibitem{O:OstaniZdrav}
{Google Play}. \#OstaniZdrav;.
\newblock
  \url{https://play.google.com/store/apps/details?id=si.gov.ostanizdrav}.

\bibitem{O:Radar:COVID}
{Google Play}. Radar COVID;.
\newblock
  \url{https://play.google.com/store/apps/details?id=es.gob.radarcovid}.

\bibitem{O:Swiss:Covid}
{Google Play}. SwissCovid;.
\newblock
  \url{https://play.google.com/store/apps/details?id=ch.admin.bag.dp3t}.

\bibitem{O:NHS:COVID:19}
{Google Play}. NHS COVID-19;.
\newblock
  \url{https://play.google.com/store/apps/details?id=uk.nhs.covid19.production}.

\bibitem{Playstore}
{Google Play}. Playstore;.
\newblock \url{https://play.google.com/store}.

\bibitem{O:EU:All:Apps}
{Europa}. Mobile contact tracing apps in EU Member States;.
\newblock
  \url{https://ec.europa.eu/info/live-work-travel-eu/coronavirus-response/travel-during-coronavirus-pandemic/mobile-contact-tracing-apps-eu-member-states_en}.

\bibitem{Androtomist}
Kouliaridis V, Kambourakis G, Geneiatakis D, Potha N.
\newblock Two Anatomists Are Better than One - Dual-Level Android Malware
  Detection.
\newblock Symmetry. 2020;12(7):1128.

\bibitem{MobSF}
{Mobile Security Framework}. Mobile Security Framework (MobSF);.
\newblock \url{https://github.com/MobSF/Mobile-Security-Framework-MobSF}.

\bibitem{O:Ostorlab}
{Ostorlab}. Ostorlab;.
\newblock \url{https://www.ostorlab.co/}.

\bibitem{O:OWASP:guide}
{OWASP}. OWASP Mobile Security Testing Guide;.
\newblock \url{https://owasp.org/www-project-mobile-security-testing-guide/}.

\bibitem{NvidiaShield}
{Nvidia}. Nvidia Shield Tablet;.
\newblock \url{https://www.nvidia.com/en-us/shield/tablet/}.

\bibitem{Apktool}
{Connor Tumbleson and Ryszard Wiśniewski}. Apktool;.
\newblock \url{https://ibotpeaches.github.io/Apktool/}.

\bibitem{API}
{Android Developers}. API reference;.
\newblock \url{https://developer.android.com/reference}.

\bibitem{qualisssllabs}
{Qualis SSL Labs}. SSL Server Test free online service;.
\newblock \url{https://www.ssllabs.com/ssltest/index.html}.

\bibitem{qualisssllabstestguide}
{Qualis SSL Labs}. SSL Server Rating Guide;.
\newblock
  \url{https://github.com/ssllabs/research/wiki/SSL-Server-Rating-Guide}.

\bibitem{policy}
{Google Support}. COVID-19 Exposure Notifications System;.
\newblock \url{https://support.google.com/android/answer/9888358?hl=en}.

\bibitem{D3PTManifest}
{DP-3T}. Decentralized Privacy-Preserving Proximity Tracing -
  AndroidManifest.xml;.
\newblock
  \url{https://github.com/DP-3T/dp3t-app-android/blob/develop/app/src/main/AndroidManifest.xml}.

\bibitem{ENS-2}
{Android Developers}. Exposure Notifications API;.
\newblock
  \url{https://developers.google.com/android/exposure-notifications/exposure-notifications-api}.

\bibitem{locationExplained}
{Google Support}. About the Exposure Notifications System and Android location
  settings;.
\newblock \url{https://support.google.com/android/answer/9930236}.

\bibitem{policy3}
{Dave Burke}. An update on Exposure Notifications;.
\newblock
  \url{https://blog.google/inside-google/company-announcements/update-exposure-notifications/}.

\bibitem{bbc}
{BBC News}. NHS Covid-19 app update blocked for breaking Apple and Google's
  rules;.
\newblock \url{https://www.bbc.com/news/technology-56713017}.

\bibitem{O:JADX}
{JADX}. skylot/jadx: Dex to Java decompiler - GitHub;.
\newblock \url{https://github.com/skylot/jadx}.

\bibitem{A:Ostorlab:1:Mabo}
Mabo T, Swar B, Aghili S.
\newblock A Vulnerability Study of Mhealth Chronic Disease Management {(CDM)}
  Applications (apps).
\newblock In: Rocha {\'{A}}, Adeli H, Reis LP, Costanzo S, editors. Trends and
  Advances in Information Systems and Technologies - Volume 1 [WorldCIST'18,
  Naples, Italy, March 27-29, 2018]. vol. 745 of Advances in Intelligent
  Systems and Computing. Springer; 2018. p. 587--598.

\bibitem{A:Ostorlab:2:Imai}
Imai H, Kanaoka A.
\newblock Chronological Analysis of Source Code Reuse Impact on Android
  Application Security.
\newblock J Inf Process. 2019;27:603--612.
\newblock doi:{10.2197/ipsjjip.27.603}.

\bibitem{Chatzoglou-2021}
Chatzoglou E, Kambourakis G, Kouliaridis V.
\newblock A Multi-Tier Security Analysis of Official Car Management Apps for
  Android.
\newblock Future Internet. 2021;13(3):58.
\newblock doi:{10.3390/fi13030058}.

\bibitem{O:NIST:SHA1:MD5}
{NIST}. Research Results on SHA-1 Collisions;.
\newblock
  \url{https://csrc.nist.gov/News/2017/Research-Results-on-SHA-1-Collisions}.

\bibitem{Janus:CVE}
{NIST}. Janus CVE;.
\newblock \url{https://nvd.nist.gov/vuln/detail/CVE-2017-13156}.

\bibitem{Android:Net:Config}
{Android Developers}. Network security configuration;.
\newblock
  \url{https://developer.android.com/training/articles/security-config}.

\bibitem{O:CWE:Mitre}
{MITRE ATT\&CK}. 2020 CWE Top 25 Most Dangerous Software Weaknesses - CWE
  Mitre;.
\newblock \url{https://cwe.mitre.org/top25/archive/2020/2020_cwe_top25.html}.

\bibitem{O:Dev:Android:Security:Tips}
{Android Developers}. Google Official Developers Android Webpage - Security
  Tips;.
\newblock \url{https://developer.android.com/training/articles/security-tips}.

\bibitem{O:Exodus:Privacy}
{Exodus}. Exodus Privacy;.
\newblock \url{https://exodus-privacy.eu.org/en/}.

\bibitem{A:Trackers:Razaghpanah}
Razaghpanah A, Nithyanand R, Vallina{-}Rodriguez N, Sundaresan S, Allman M,
  Kreibich C, et~al.
\newblock Apps, Trackers, Privacy, and Regulators: {A} Global Study of the
  Mobile Tracking Ecosystem.
\newblock In: 25th Annual Network and Distributed System Security Symposium,
  {NDSS} 2018, San Diego, California, USA, February 18-21, 2018. The Internet
  Society; 2018.

\bibitem{A:Trackers:2:Vallina}
Vallina{-}Rodriguez N, Sundaresan S, Razaghpanah A, Nithyanand R, Allman M,
  Kreibich C, et~al.
\newblock Tracking the Trackers: Towards Understanding the Mobile Advertising
  and Tracking Ecosystem.
\newblock CoRR. 2016;abs/1609.07190.

\bibitem{A:Trackers3:Liu}
Liu X, Liu J, Zhu S, Wang W, Zhang X.
\newblock Privacy Risk Analysis and Mitigation of Analytics Libraries in the
  Android Ecosystem.
\newblock {IEEE} Trans Mob Comput. 2020;19(5):1184--1199.
\newblock doi:{10.1109/TMC.2019.2903186}.

\bibitem{O:Firebase}
{Firebase}. Firebase;.
\newblock \url{https://firebase.google.com/}.

\bibitem{O:Appcenter:Crashes}
{Microsoft}. Appcenter Android Crashes;.
\newblock \url{https://docs.microsoft.com/en-us/appcenter/sdk/crashes/android}.

\bibitem{O:MS:Analytics}
{Microsoft}. Microsoft Visual Studio App Center Analytics;.
\newblock \url{https://docs.microsoft.com/en-us/appcenter/analytics/}.

\bibitem{O:Matomo}
{Matomo}. Matomo | Google Analytics alternative analytics tracker;.
\newblock \url{https://matomo.org/}.

\bibitem{O:Bugsnag}
{Bugsnag}. Bugsnag: Error Monitoring \& App Stability Management tracker;.
\newblock \url{https://www.bugsnag.com/}.

\bibitem{O:Android:Manifest}
{Android Developers}. App Manifest Overview | Android Developers;.
\newblock
  \url{https://developer.android.com/guide/topics/manifest/manifest-intro}.

\bibitem{O:Intent:Malicious}
{Therese Mendoza}. Compromising Android Applications with Intent Manipulation;.
\newblock
  \url{https://www.trustwave.com/en-us/resources/blogs/spiderlabs-blog/compromising-android-applications-with-intent-manipulation/}.

\bibitem{O:Dev:Android:NDK}
{Android Developers}. Google Official Developers Android Webpage - Android
  NDK;.
\newblock \url{https://developer.android.com/ndk}.

\bibitem{A:Shared:Libraries:Tegawend}
Li L, Bissyand{\'{e}} TF, Klein J, Traon YL.
\newblock An Investigation into the Use of Common Libraries in Android Apps.
\newblock In: {IEEE} 23rd International Conference on Software Analysis,
  Evolution, and Reengineering, {SANER} 2016, Suita, Osaka, Japan, March 14-18,
  2016 - Volume 1. {IEEE} Computer Society; 2016. p. 403--414.

\bibitem{A:Shared:Libraries:Taylor}
Taylor VF, Beresford AR, Martinovic I.
\newblock Intra-Library Collusion: {A} Potential Privacy Nightmare on
  Smartphones.
\newblock CoRR. 2017;abs/1708.03520.

\bibitem{Backes-2016}
Backes M, Bugiel S, Derr E.
\newblock Reliable Third-Party Library Detection in Android and its Security
  Applications.
\newblock In: Weippl ER, Katzenbeisser S, Kruegel C, Myers AC, Halevi S,
  editors. Proceedings of the 2016 {ACM} {SIGSAC} Conference on Computer and
  Communications Security, Vienna, Austria, October 24-28, 2016. {ACM}; 2016.
  p. 356--367.

\bibitem{O:Shared:Libraries:RELRO}
{Julian Cohen}. RELRO: RELocation Read-Only;.
\newblock
  \url{https://medium.com/@HockeyInJune/relro-relocation-read-only-c8d0933faef3}.

\bibitem{A:Third:Party:Libraries:Derr:2017}
Derr E, Bugiel S, Fahl S, Acar Y, Backes M.
\newblock Keep me Updated: An Empirical Study of Third-Party Library
  Updatability on Android.
\newblock In: Thuraisingham BM, Evans D, Malkin T, Xu D, editors. Proceedings
  of the 2017 {ACM} {SIGSAC} Conference on Computer and Communications
  Security, {CCS} 2017, Dallas, TX, USA, October 30 - November 03, 2017. {ACM};
  2017. p. 2187--2200.

\bibitem{A:Third:Party:Libraries:Salza:2018}
Salza P, Palomba F, Nucci DD, D'Uva C, Lucia AD, Ferrucci F.
\newblock Do developers update third-party libraries in mobile apps?
\newblock In: Khomh F, Roy CK, Siegmund J, editors. Proceedings of the 26th
  Conference on Program Comprehension, {ICPC} 2018, Gothenburg, Sweden, May
  27-28, 2018. {ACM}; 2018. p. 255--265.

\bibitem{O:nvd:nist}
{NIST}. NVD NIST;.
\newblock \url{https://nvd.nist.gov/}.

\bibitem{O:sqlite}
{SQLite}. SQLite;.
\newblock \url{https://www.sqlite.org/index.html}.

\bibitem{O:openssl}
{OpenSSL}. OpenSSL;.
\newblock \url{https://www.openssl.org/}.

\bibitem{O:libjpeg}
{libjpeg}. libjpeg;.
\newblock \url{http://libjpeg.sourceforge.net/}.

\end{thebibliography}
\end{document}